\documentclass{autart}

\usepackage{amsmath, amssymb}
\usepackage{algpseudocode} 
\usepackage{algorithm} 
\usepackage{algorithmicx} 
\usepackage{graphicx}
\graphicspath{{fig/}{../fig/}}
\usepackage{cuted}
\usepackage{multicol}
\usepackage[latin1]{inputenc}
\usepackage{lipsum}  
\usepackage{textcomp}
\def\BibTeX{{\rm B\kern-.05em{\sc i\kern-.025em b}\kern-.08em
    T\kern-.1667em\lower.7ex\hbox{E}\kern-.125emX}}

\usepackage{array}
\usepackage{makecell}
\newcommand{\PreserveBackslash}[1]{\let\temp=\\#1\let\\=\temp}
\newcolumntype{C}[1]{>{\PreserveBackslash\centering}p{#1}}
\newcolumntype{R}[1]{>{\PreserveBackslash\raggedleft}p{#1}}
\newcolumntype{L}[1]{>{\PreserveBackslash\raggedright}p{#1}}

\usepackage{color}

\providecommand{\com[2]}{\begin{tt}[#1: #2]\end{tt}}

\newcommand{\beq}{\begin{equation}}
\newcommand{\eeq}{\end{equation}}
\newcommand{\beqa}{\begin{eqnarray}}
\newcommand{\eeqa}{\end{eqnarray}}
\newcommand{\beqan}{\begin{eqnarray*}}
\newcommand{\eeqan}{\end{eqnarray*}}
\newcommand{\bef}{\begin{figure}}
\newcommand{\enf}{\end{figure}}

\definecolor{magenta}{cmyk}{0.1, 1, 0, 0}
\definecolor{greenedu}{cmyk}{1, 0, 1, 0.1}
\definecolor{cyanedu}{cmyk}{1, 0, 0, 0.1}
\definecolor{brown}{cmyk}{0.2, 0.63, 0.84, 0.2}
\definecolor{BZcolor}{cmyk}{0, 0.35, 1, 0.2}

\newcommand{\bi}{\begin{itemize}}
\newcommand{\ei}{\end{itemize}}
\newcommand{\bc}{\begin{center}}
\newcommand{\ec}{\end{center}}
\newcommand{\ba}{\begin{array}}
\newcommand{\ea}{\end{array}}
\newcommand{\be}{\begin{equation}}
\newcommand{\ee}{\end{equation}}
\newcommand{\beno}{\begin{equation*}}
\newcommand{\eeno}{\end{equation*}}
\newcommand{\beqna}{\begin{eqnarray}}
\newcommand{\eeqna}{\end{eqnarray}}
\newcommand{\bd}{\begin{displaymath}}
\newcommand{\ed}{\end{displaymath}}
\newcommand{\beqnd}{\begin{eqnarray*}}
\newcommand{\eeqnd}{\end{eqnarray*}}

\renewcommand{\ni}{\noindent}

\newcommand{\cqfd}{\hfill \rule{2mm}{2mm}\medbreak\indent}

\newcommand{\emp}[1]{$\text{EMP}_{#1}$}

\newenvironment{proof}{\noindent\textbf{Proof.} }{\hfill $\blacksquare$ \par}
\newtheorem{theorem}{\bf Theorem}[section]
\newtheorem{lemma}{\bf Lemma}[section]

\newtheorem{proposition}{\bf Proposition}[section]
\newtheorem{corollary}{\bf Corollary}[section]

\newtheorem{definition}{\bf Definition}

\definecolor{red}{rgb}{1,0,0}

\definecolor{blu}{rgb}{0,0,1}

\definecolor{gre}{rgb}{0,0.7,0.3}

\definecolor{bz}{rgb}{1,.7,0}

\begin{document}

\begin{frontmatter}

\title{A graph decomposition approach for the synthesis of EMPs in directed acyclic networks\thanksref{footnoteinfo} }

\author[Eduardo]{Eduardo Mapurunga}$^c$\ead{mapurunga@ufc.br},    
\author[Michel]{Michel Gevers}\ead{michel.gevers@uclouvain.be},               
\author[BZ]{Alexandre Sanfelici Bazanella}\ead{bazanella@ufrgs.br}  

\address[Eduardo]{Universidade Federal do Cear{\'a}, Itapaj{\'e}, Brazil}                                                
\address[Michel]{UCLouvain Institute of Information and Communication Technologies, Electronics and Applied Mathematics (ICTEAM), Louvain la Neuve, Belgium}            
\address[BZ]{Data Driven Control Group, Department of Automation and Energy, 
Universidade Federal do Rio Grande do Sul, Porto Alegre-RS, Brazil}

\thanks[footnoteinfo]{
This work  was supported in part by Conselho Nacional de Desenvolvimento Cient{\' i}fico e Tecnol{\' o}gico (CNPq), through a post-doctoral fellowship (177257/2023-7) to the first author, and through the universal project 404615/2023-6, and by Wallonie-Bruxelles International (WBI).}
%


\begin{abstract}
This paper considers the identifiability of dynamical networks whose topology is a Directed Acyclic Graph (DAG), in the context where not all nodes are excited and not all nodes are measured. 
The objective is to construct a valid and parsimonious 
 Excitation and Measurement Pattern (EMP), i.e. a selection of excited and measured nodes, that allows to identify the whole network with a small number of excitations and measurements. We show first that a DAG can be decomposed into a set of trees and parallel paths networks. We propose an algorithm that achieves this decomposition and computes an EMP that combines EMPs for the individual trees and parallel paths networks.
				We formally show that this algorithm yields an EMP capable of identifying the whole network. 

\end{abstract}

\end{frontmatter}

\section{Introduction}
\label{intro}

Dynamical networks are present in many aspects of everyday life, such as communication networks, distribution networks, epidemic propagation models, chemical reaction networks, genetic networks, and many more. As a consequence, a very large amount of papers have been devoted for a long time to various questions related to dynamical networks. For example, network models have been used to study the propagation of epidemics in \cite{Pare-Beck-Basar:2020}, for the modeling of distribution grids in power systems in \cite{Ardakanianetall:2019}, for applications in the oil industry in \cite{Munaro:2015}. Identifiability of weighted consensus networks have been studied in \cite{NabaviChakrabortty:2016}.

The structure of dynamical networks is typically described by a graph that connects the nodes of the network to its other nodes, via links that are called edges. 
Central among the problems raised by dynamical networks have been the construction, from data, of the graph that describes the topology of the network, and the identification of the components that make up the edges of the graph, when it is assumed that the topology is known. 

An important contribution for the problem of topology reconstruction can be found in the seminal papers \cite{Materassi-Innocenti:2010} and \cite{Materassi-Salapaka:2012}. In \cite{MattaSayed:2019}, which builds on these two papers, the problem of topology reconstruction is called network tomography. All these papers are 
 based on the detection of causality and feedback between time-series, using the early works on Granger causality \cite{Granger:63}, \cite{Gevers-Anderson:82}, \cite{Caines:88}. An entirely different approach to the topology reconstruction problem is taken in \cite{Monika-Petreczky-Camlibel:2024}.
 Their approach does not rely on detecting causality between multiple time-series, as in the Granger causality approach; rather they consider deterministic networks of linear dynamical systems. They provide necessary and sufficient conditions under which the network can be reconstructed from data, and they provide a framework for this reconstruction based on constrained Lyapunov equations.
 
Most of the work on identification of dynamical networks has been directed to the problem of identifiability of its components (i.e. its edges) assuming that the topology is known. The network is typically described by a network matrix, whose elements represent the contents of the links, if any, between any couple of nodes. Essentially, the work of the last decade has considered that links between the nodes are rational transfer functions.

When it comes to the identification of these transfer functions at the edges of the network, 
starting from the seminal paper \cite{vandenhof-dankers-heuberger-etal-identification-2013},
most of the work has been directed to the problem of identifiability, that is, whether or
not it is possible to determine these transfer functions uniquely from the available input/output (I/O) data. 
Actual identification methods have not been so much studied, although there are results 
concerning the consistency and efficiency of the estimates for some methods under the 
assumption of identifiability \cite{vandenhof-dankers-heuberger-etal-identification-2013,rajagopal-ramaswamy-vandenhof-learning-2021,gevers-bazanella-dasilva-practical-2018,ramaswamy-hof-local-2021}. 
It has been found that identifiability is essentially determined by the properties of the underlying directed graph (called digraph),  
and by which I/O pairs can be directly obtained from the data. 
Accordingly, conditions for identifiability are given in terms of two aspects: the (given) topology of the network, and the selection of excited and/or measured nodes. Such conditions have been studied under 
different assumptions, resulting in various fundamental results 
\cite{bazanella-gevers-hendrickx-etal-identifiability-2017,hendrickx-gevers-bazanella-identifiability-2019,MaterassiSalapaka:2020,JahandariMaterassi:2022,van_waarde_necessary_2019,van_waarde_identifiability_2018,cheng-shi-vandenhof-allocation-2021,weerts-vandenhof-dankers-identifiability-2018}, but mostly under the hypothesis that either
all nodes are excited or all nodes are measured.


The more general - and much more relevant - case in which one is not constrained to either
excite all nodes or observe all nodes is much more involved and has been addressed for the first time in \cite{bazanella-gevers-hendrickx-network-2019},
where a set of necessary conditions for identifiability has been presented. So, now the problem becomes one of finding
specific combinations of excited nodes and of measured nodes
- so-called excitation and measurement
patterns, or {\em EMP}s - such that the network edges can be identified. These are then called {\em valid EMP}s.  
Necessary and sufficient conditions that must be satisfied by any {\em valid EMP} for specific classes of graphs 
- trees, single loops and Parallel Paths Networks - have been given in \cite{bazanella-gevers-hendrickx-network-2019,mapurunga-gevers-bazanella-necessary-2022,mapurunga-gevers-bazanella-CDC2024}. On the other hand, it is rather easy to realize that in any valid 
EMP all sources\footnote{Nodes with no incoming edges} must be excited and all sinks\footnote{Nodes with no outcoming edges}
must be measured, but far from easy is the realization that there are other types of nodes that also must be excited
(these have been called dources) or measured (called dinks),
as proven in \cite{mapurunga-gevers-baza-DDJournal-2024}.

Two other approaches to the design of valid EMPs, with different objectives than parsimony, have been developed recently. One is to design an EMP such that some measure of the accuracy of the estimated edge transfer functions is minimized \cite{mapurunga-optimal-2021,mapurunga-bazanella-optimal-excitation-2023}. Another is to design  a valid EMP that minimizes the cost of excitation and measurement of the different nodes \cite{JahandariMaterassi:2022,Bazanella&Gevers&Mapurunga:CDC2025}. Finally, we should mention that, whatever the objective, testing whether a given EMP is valid has now been made straightforward (though computationally intensive) thanks to a test developed in \cite{legat-hendrickx-pathbased-2021a}.

%

In this paper we address the classical situation  where sparsity of the EMP is desirable.
Clearly, a valid EMP can be obtained by just exciting and measuring all nodes, but 
what we look for is a parsimonious EMP, that is one in
which few measurements and excitations are applied. Except for the specific classes of network topologies described above (trees, loops and Parallel Paths Networks), for which minimal EMPs have been obtained, the design of valid (not necessarily minimal) EMPs for general topologies is still a difficult task for networks with general topologies.

We present an algorithm  for the systematic design of EMPs for Directed Acyclic Graphs (DAGs), which are networks without
loops. Preliminary results on the construction of EMPs for DAGs (but without a systematic algorithm based on the decomposition described below) were presented in \cite{mapurunga-gevers-baza-CDC2022}. 

{The new algorithm presented in this paper is based on the complete decomposition of the DAG into two building blocks: Induced Trees (IT) and Parallel Paths Networks (PPN). 
We first show that any node in a network is the source of a uniquely defined tree, which we call the \textbf{Induced Tree (IT)} of that node. An IT is a tree, for which we know how to construct a minimal EMP \cite{bazanella-gevers-hendrickx-network-2019}.   A \textbf{Parallel Paths Network (PPN)} is a network that contains several paths that link a single source to a single sink. Since any DAG can be completely decomposed into  subsets of ITs and PPNs, we can then apply our earlier results \cite{bazanella-gevers-hendrickx-network-2019,mapurunga-gevers-bazanella-CDC2024} to design valid EMPs for these ITs and PPNs. Actually, the results of \cite{bazanella-gevers-hendrickx-network-2019,mapurunga-gevers-bazanella-CDC2024} show how to construct minimal EMPs for these ITs and PPNs.}

{However, nothing guarantees that when we design valid (or minimal) EMPs for subsets of networks, they remain valid for the combined network. Conditions for this property to hold have been presented in \cite{mapurunga-gevers-bazanella-CDC2024}. An essential element of our contribution is to show that the DAG decomposition algorithm that we develop is  constructed in such a way that the valid EMPs designed for these subsets of ITs and PPNs taken in isolation remain valid when these subsets are combined to form the whole DAG, i.e. that the conditions of \cite{mapurunga-gevers-bazanella-CDC2024} are satisfied.}

{As stated above, the algortihm presented in this paper aims at designing a valid EMP for a DAG network while keeping the cardinality of this EMP low, but this cardinality will typically not be minimal. Computing a minimal EMP for a DAG is beyond the scope of this paper, and is probably without reach. An exhaustive search appears to be the only way to obtain with certainty a minimal EMP, but the computation time would be staggering even for moderate size networks. However, two features in the algorithm of this paper tend to keep the cardinality low. First,  the algorithm is based on designing a minimal EMP for each IT of the decomposition. Given that there is a large degree of freedom to select a minimal EMP for a tree, the second feature is that, for each IT, the algorithm selects a particular EMP that insures that the PPNs that connect with this IT are identifiable. This is achieved by steps 11 and 12 of the algorithm. As a result, all PPNs become identifiable without the need for additional excitations and measurements. }

{The implementation of the algorithm developed in Section~\ref{sec:decalg} is simple, and in the end it is based on designing minimal EMPs for the ITs only. But to prove that it produces a valid EMP requires a lot of results on ITs, PPNs and their interaction, and on the identifiablity requirements when subgraphs equipped with valid EMPs are combined. Some of these results were produced in our earlier papers \cite{bazanella-gevers-hendrickx-network-2019,mapurunga-gevers-bazanella-CDC2024}, but the notion of Induced Tree, which is crucial for the decomposition of the DAG into ITs and PPNs,  is novel, as well as the results on the  properties of ITs and their interactions with PPNs. Finally, the theorems of Section~\ref{sec:decalg}, which validate the algorithm, are a major new contribution.}


We have coded this algorithm in Python, and we provide an example
of moderate size (25 nodes) to illustrate the parsimonious nature of the resulting EMP. 
An alternative procedure for the construction of EMPs for DAGs, based on a decomposition of the DAG into disjoint trees, has been proposed in \cite{cheng-shi-lestas-vandenhof-DAGpartial-2025}.


The paper is organized as follows.
				In Section \ref{sec:defs} we describe the class of dynamical networks for which our results are derived, and we recall some basic properties of such networks and of the EMPs. In Section \ref{sec:structures} we present two basic structures, namely Parallel Paths Networks (PPN) and Induced Trees (IT), as well as  identifiability results for these two specific  topologies treated in isolation. Our focus on these two specific structures will become clear in 
Section \ref{sec:decDAGs} where we show that a Directed Acyclic Graph (DAG) can be decomposed into two subdigraphs  composed  of  ITs and PPNs. In Section \ref{sec:decalg} an  algorithm is presented that provides a valid EMP for DAGs; it is based on this decomposition scheme of a DAG into ITs  and PPNs. A simple tutorial example is presented in Section~\ref{sec:tutorial} to explain the workings of the algorithm. 				In Section \ref{sec:example} we apply the algorithm to  an example of moderate size. 
				The conclusions  are presented in Section \ref{sec:conclu}.

\section{Definitions, Notations and Preliminaries}
\label{sec:defs}

In this section, we introduce the dynamical networks that we are dealing with, and we introduce the notations used for these networks and their associated digraphs. We recall the necessary conditions for identifiability that were derived in \cite{bazanella-gevers-hendrickx-network-2019} and \cite{mapurunga-gevers-baza-DDJournal-2024}, and we introduce the concept of  a valid Excitation and Measurement Pattern, namely a choice of excited nodes and measured nodes that guarantees identifiability. 

We consider dynamical networks composed of $n$ nodes (or vertices) which represent internal scalar signals $\left\lbrace w_k(t) \right\rbrace$ for $k \in  \{1, 2, \dots, n\}$.
These nodes are interconnected by discrete time transfer functions, represented by edges, which are entries of a \emph{network matrix} $G(z)$.
The dynamics of the network are given by the following equations:
\begin{subequations}
	\begin{align}
	w(t) &= G(z)w(t) + Br(t), \label{eq:dynet1} \\
	y(t) &= Cw(t), \label{eq:dynet2}
	\end{align}
\end{subequations}
where $w(t) \in \mathbb{R}^n$ is the node vector, $r(t) \in \mathbb{R}^m$ is the input vector, and $y(t) \in \mathbb{R}^p$ is the set of measured nodes, considered as the output vector of the network.
The matrix $B \in \mathbb{Z}_2^{n \times m}$, where $\mathbb{Z}_2 \triangleq \{0, 1\}$, 
is a binary selection matrix with a single $1$ and $n-1$ zeros  in each column; it selects the inputs affecting the nodes of the network.
Similarly, $C \in \mathbb{Z}_2^{p \times n}$ is a matrix with a single $1$ and $n-1$ zeros in each row that selects which nodes are measured. 

 To each network matrix $G(z)$ we can associate a directed graph (called digraph) $\mathcal{G}$ defined by the tuple $(\mathcal{V}, \mathcal{E})$, where $\mathcal{V}$ is the set of vertices and $\mathcal{E} \subseteq \mathcal{V} \times \mathcal{V}$ is the set of edges. The digraph $\mathcal{G}$ defines the topology of the network.
	The nodes are related by transfer functions. {We  denote by $(j,i)$ an edge  that goes from node $j$ to node $i$; } node $j$ is then called an in-neighbor of node $i$, while node $i$ is an out-neighbor of node $j$.
Thus, there is an edge $(j,i)$ associated with the transfer function $G_{ij}(z)$ of the network matrix only if $G_{ij}(z)$ is nonzero. 	{We say that a digraph is \emph{connected} if every node can be reached from every other  node in the network when the edges are treated as undirected.}			

For the digraph $\mathcal{G}$ associated to the network matrix $G(z)$ we introduce the following notations.
\begin{itemize}
	\item $\mathcal{V}$ - the set of all $n$ nodes;
	\item $\mathcal{B}$ - the set of excited nodes, defined by $B$ in (\ref{eq:dynet1});
	\item $\mathcal{C}$ - the set of measured nodes, defined by $C$ in (\ref{eq:dynet2});
	\item $\mathcal{F}$ - the set of sources: nodes with no incoming edges;
	\item $\mathcal{D}_F$ - the set of dources: see Definition \ref{def:dource} below;
	\item $\mathcal{S}$ - the set of sinks: nodes with no outgoing edges;	
	\item $\mathcal{D}_S$ - the set of dinks: see Definition \ref{def:dink} below;
	\item $\mathcal{I}$ - the set of internal nodes, i.e. nodes that are neither a source nor a sink: $\mathcal{I} \triangleq \mathcal{V}\backslash(\mathcal{F}\cup\mathcal{S})$;
	\item $\mathcal{N}_j^-$ -  the set of in-neighbors  of node $j$;
	\item $\mathcal{N}_j^+$ - the set of out-neighbors of node $j$.
\end{itemize}
We recall the definitions of dources and dinks \cite{mapurunga-gevers-baza-DDJournal-2024}.
\begin{definition}
				A node $j$ is called a \textbf{dource} if it has at least one out-neighbor to which all  its in-neighbors have a directed edge.				\label{def:dource}
\end{definition}

\begin{definition}
				A node $j$ is called a \textbf{dink} if it has at least one in-neighbor that has a directed edge to all  its
  out-neighbors.		\label{def:dink}
\end{definition}
We observe that a node can be both a dource and a dink, and we refer to \cite{mapurunga-gevers-baza-DDJournal-2024} for details.

{\ni \bf Assumptions on the network matrix $G(z)$}\\
Throughout the paper, we shall make the following assumptions on the network matrix:
\begin{itemize}
	\item the diagonal elements are zero and all other elements are  proper;
	\item $(I - G(z))^{-1}$ is proper and all its elements are stable.
\end{itemize}

One can represent the dynamical network in (\ref{eq:dynet1})-(\ref{eq:dynet2}) as an input-output model as follows
\begin{align}
	y(t) = M(z) r(t),\; \text{with}\; M(z) \triangleq C T(z) B
	\label{eq:IOdynet}
\end{align}
where
\begin{align}
	T(z) \triangleq (I - G(z))^{-1}. \label{eq:Tdef}
\end{align}
Observe that the matrix $T(z)$ is generically nonsingular by construction. 

In analyzing the identifiability of the network matrix, it is assumed that the input-output model $M(z)$ is known; the identification of $M(z)$ from  input-output (IO) data $\lbrace y(t), r(t) \rbrace$ is a standard  identification problem, provided the input signal $r(t)$ is sufficiently rich. The question of identifiability of the network is then whether the network matrix $G(z)$ can be fully recovered from the  known transfer matrix $M(z)$. We now give a formal definition of generic identifiability of the network matrix from the data $\lbrace y(t), r(t) \rbrace$ and from the digraph structure.

{
\begin{definition}
(\cite{hendrickx-gevers-bazanella-identifiability-2019}) Let $G(z)$ be a network matrix  and let $G(P, z)$ be any rational transfer matrix parametrization  consistent with the  digraph associated with $G(z)$, with the parameter vector $P \in 
\mathbb{R}^{N}$. Then  $G(z)$ is generically identifiable from excitation signals applied to $\mathcal{B}$ and measurements made at $\mathcal{C}$ if
\[
					C [I - G(P, z)]^{-1} B = C[I - \tilde{G}(z)]^{-1} B \implies G(P, z) = \tilde{G}(z),
	\]
for all parameters $P$ except possibly those lying on a zero measure set in $\mathbb{R}^{N}$, where $\tilde{G}(z)$ is any network matrix consistent with the digraph.
\end{definition}
}


In this paper, we discuss generic 
identifiability  in terms of which nodes must be excited and/or measured in order to guarantee identifiability of the network. Thus, we do not assume that either all nodes are excited or all nodes are measured, which was the common assumption until the publication of \cite{bazanella-gevers-hendrickx-network-2019}. Instead, we examine which constraints must be imposed on an EMP in order for it to be valid. 
The following Proposition combines necessary conditions for generic identifiability of a network; 
it combines Theorem III.1, Corollary III.1 from \cite{bazanella-gevers-hendrickx-network-2019} with Theorem III.1 from \cite{mapurunga-gevers-baza-DDJournal-2024}. 

\begin{proposition}\label{corol1}
The network matrix $G(z)$ is generically identifiable only if $\mathcal{B}, \mathcal{C} \neq \emptyset$; $\mathcal{F}, \mathcal{D}_F \subset \mathcal{B}$; $\mathcal{S}, \mathcal{D}_S \subset \mathcal{C}$  and $\mathcal{B} \cup \mathcal{C} = \mathcal{V}$.\label{cor:fromBandC}
\end{proposition}

Finding conditions that guarantee generic  identifiability of a given network  is equivalent to constructing an Excitation and Measurement Pattern (EMP)  that guarantees identifiability; such EMP is then called a \emph{valid EMP}.
The concept of EMP and of valid EMP, which led  to the concept of minimal EMP, was introduced in \cite{mapurunga-optimal-2021}. They are defined in the following. 
\begin{definition}
    A pair of selection matrices $B$ and $C$, with its corresponding pair of node sets $\mathcal{B}$ and $\mathcal{C}$,  is called an \textbf{excitation and measurement pattern} - EMP for short.
  An EMP is said to be \textbf{valid} for the network (\ref{eq:dynet1})-(\ref{eq:dynet2})
  if it is such that this network is generically identifiable.
  Let $\nu = |\mathcal{B}| + |\mathcal{C}|$ \footnote{$|\cdot|$ - Denotes the cardinality of a set.} be the cardinality of an EMP.
  A given EMP is said to be  \textbf{minimal} if it is valid and there is no other valid EMP with smaller cardinality.
  \label{def:EMP}
\end{definition}

The following result establishes a lower and an upper bound for the cardinality of a valid EMP for any network.
\begin{lemma}\label{cardinality}
The cardinality of a minimal EMP for the identification of a dynamical network with $n$ nodes is at least equal to $n$ and at most equal to $2n-f-s$, where $f$ is the number of sources and $s$ the number of sinks.
\end{lemma}
\begin{proof} 
				See \cite{mapurunga-gevers-baza-DDJournal-2024}. 
\end{proof}

Proposition~\ref{corol1} states that for an EMP to be valid, it must contain at least one excitation and one measurement, all sources and dources must be excited and all sinks and dinks must be measured, and every other node must be either excited or measured. 
In Theorems V.1 and V.2 from \cite{mapurunga-gevers-baza-DDJournal-2024} it has been shown that sources and dources are the only types of nodes that always require excitation, while sinks and dinks are the only ones that always require measurement. 

From now on, we drop the arguments $z$ and $t$ used in (\ref{eq:dynet1})-(\ref{eq:dynet2}) whenever there is no risk of confusion.

\section{Trees, parallel paths networks and induced trees}
\label{sec:structures}

In this paper we are interested in obtaining valid EMPs for networks whose topologies are associated  with directed acyclic digraphs, referred to as DAGs\footnote{For simplicity, we shall in the future just refer to a DAG rather than a network that has the topology of a DAG.}. One of the main results of this paper will be to show that a DAG can be completely covered by  a set of so-called Induced Trees (IT) and Parallel Paths Networks (PPN), 
which are special cases of DAGs.
In this section, we recall  necessary and sufficient conditions for the generic identifiability of directed trees and of parallel paths networks, established in \cite{hendrickx-gevers-bazanella-identifiability-2019} and \cite{mapurunga-gevers-bazanella-CDC2024}, respectively, and we introduce the concept of induced tree. 
The identifiability conditions for directed trees and PPNs will serve as a backbone for designing valid EMPs for more general DAGs.

A DAG is the most general class of acyclic network, which encompasses several acyclic topologies. 
Directed Acyclic Graphs are defined as follows.
\begin{definition}\cite{deoGraphTheoryApplications1974}
				A directed acyclic graph (DAG) is a directed graph that has no cycles.
    \label{def:DAG}
\end{definition}
A directed tree can be defined as follows. 


\begin{definition}
				\cite{deoGraphTheoryApplications1974}
				A directed tree is a minimally connected DAG 
	-- removal of any edge makes the directed tree to be disconnected. 
	\label{def:tree}
\end{definition}


Necessary and sufficient conditions for the identifiability of a directed tree are as follows.

\begin{theorem}
\cite{bazanella-gevers-hendrickx-network-2019}  A directed tree is generically  identifiable if and only if  
 ${\mathcal F}\subseteq {\mathcal B}$,
 ${\mathcal S}\subseteq {\mathcal C}$,
 ${\mathcal B}\cup {\mathcal C} = {\mathcal V}$.
 \label{theo:iso-trees}
\end{theorem}
Notice that there are no dources and dinks in directed trees.

Parallel paths networks (PPN) are DAGs with a specific structure. They were introduced in \cite{mapurunga-gevers-bazanella-CDC2024}, where necessary and sufficient conditions for their generic identifiabillity were presented. We briefly recall the main results of \cite{mapurunga-gevers-bazanella-CDC2024} on PPNs. 

\begin{definition}
  A \textbf{parallel paths network (PPN)}  is a network composed of a single source and a single sink. 
	There are $n_p \ge 2$ paths from the source to the sink,
	which are the only nodes common to each path.
	Every path other than a direct edge between source and sink has at least three nodes.
	\label{def:parallel}
\end{definition}

\begin{figure}
\centering
	\includegraphics[width=0.5\linewidth]{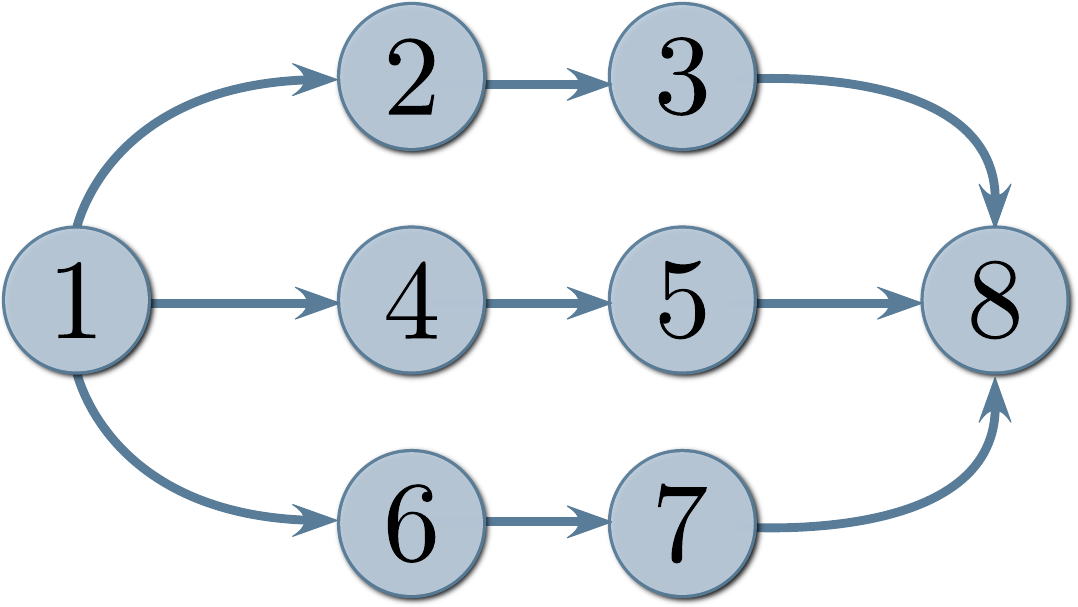} \hfill \includegraphics[width=0.4\linewidth]{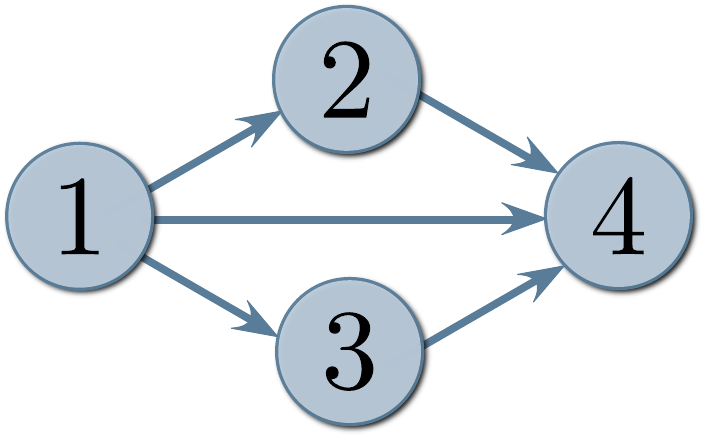}
	\caption{
									Examples of PPNs from Definition \ref{def:parallel}.
									The graph on the left depicts a PPN with three paths,
									each of which with 2 internal nodes; the one on the right 
									depicts a PPN wherein one path is a direct link from source to sink.				}
\label{fig:Ex-PPNs}
\end{figure}

				Examples of PPNs are depicted in Figure~\ref{fig:Ex-PPNs}. 
				A PPN has several paths between the source node and the sink node. 
				The main characteristic of a PPN is that each node in a path, except for the source, has only one path linking it to the sink node. 
								Notice that a PPN may have a direct edge linking the source to the sink node, as shown in the example depicted on the right of Figure~\ref{fig:Ex-PPNs}. The following theorem gives necessary and sufficient conditions for the generic identifiability of a PPN. 

\begin{theorem} \cite{mapurunga-gevers-bazanella-CDC2024}
				 Consider a parallel paths network from Definition \ref{def:parallel} with 
				one source 
				$\mathcal{F} := \{1\} $
				connected through $n_{p} \ge 2$ paths $\left(\mathcal{P}_1, \mathcal{P}_2, \dots, \mathcal{P}_{n_p} \right)$  
				to one sink $\mathcal{S} := \{L\} $.				
								Let $\mathcal{V}_{\mathcal{P}_j}$ be the set of nodes of  path $\mathcal{P}_j$. 
												Assume that the nodes have been labeled sequentially,
				such that for every path $\mathcal{P}_{j}$ there exists a path from $k'_j$ to $k''_j$ only if $k'_j < k''_j$. 
				A parallel paths network is generically identifiable if and only if 
				$\mathcal{F} \subset \mathcal{B}$, 
				$\mathcal{S} \subset \mathcal{C}$, 
				$\mathcal{B} \cup \mathcal{C} = \mathcal{V}$, and in addition
				there are $n_p - 1$ paths $\mathcal{P}_j$ for which there exist at least 
				$k_j', k_j'' \in \mathcal{V}_{\mathcal{P}_j} \setminus \{1, L\} $, such that $k_j' \le k_j''$, $k_j' \in \mathcal{B}$,  $k_j'' \in \mathcal{C}$.     
				\label{theo:generalparallel}
\end{theorem}

The following new result yields the cardinality of a minimal EMP for the identification of a PPN.
\begin{theorem} 
				 Consider a parallel paths network that has  $n$ nodes.	The cardinality of a minimal EMP for such PPN is:\\
$\bullet$ $n$ if the PPN has no paths that contain a single internal node\\
$\bullet$ $n + K_1 + K_2 -1$ if the PPN has at least one path that contains a single internal node, where $K_1$ is the number of paths that contains a single internal node, while $K_2=1$ if there is a direct link between the source and the sink, and zero otherwise.
				\label{theo:minimalEMPforPPN}
\end{theorem}
\begin{proof}
(i) Consider first that there exists no path with a single internal node. According to Theorem~\ref{theo:generalparallel} a valid EMP is defined as follows. For each of the $n_p$ paths, excite the first internal node and measure all others. This EMP is minimal since its cardinality is $n$.\\
(ii) Consider now that there are $K_1$ paths with one internal node. Two cases must be considered.
\begin{enumerate}
\item If there is a direct edge $1 \to L$, then in all other $n_p-1$ paths, the last condition of Theorem~\ref{theo:generalparallel} must be satisfied. This requires that, in the $K_1$ paths with one internal node, this node is both excited and measured. In the  $n_p-1-K_1$ remaining nodes, excite the first internal node and measure all others. By Theorem~\ref{theo:generalparallel}, this EMP is minimal  and its cardinality is $n+K_1$.
\item If there is no direct edge $1 \to L$, then one of the $K_1$ paths with one internal node can be taken as the $n_p$-path, and the number of paths with one internal node that requires both excitation and measurement is reduced by one. 
\end{enumerate}
The cardinality of a minimal EMP for these two situations is thus captured by the formula  $n + K_1 + K_2 -1$.
\end{proof}
Designing EMPs for PPNs is  a rather simple task. Consider the example depicted on the left of  Figure \ref{fig:Ex-PPNs}. 
				There are three paths, each of which with two internal nodes: $\{2, 3, 4, 5, 6, 7\}$. 
				 In order to obtain a valid EMP, one can choose two paths to which the last condition of Theorem~\ref{theo:generalparallel} are applied, say  $ 1 \to 2 \to 3 \to 8$ and $1 \to 4\to 5 \to 8$.  
				Node $1$ is a source and it must be excited, while node $8$ is a sink and it must be measured. 
				To obey the conditions of Theorem~\ref{theo:generalparallel} one must  excite nodes $\{2, 4\}$ and measure nodes $\{3, 5\}$. 
				The remaining nodes $\{6, 7\}$ can each be either excited or measured, resulting in either one of the following valid \emp{1} $=(\mathcal{B}_1 = \{1, 2, 4, 6\}, \mathcal{C}_1 = \{3, 5, 7, 8\})$, or \emp{2} $=(\mathcal{B}_1 = \{1, 2, 4, 7\}, \mathcal{C}_1 = \{3, 5, 6, 8\})$. These EMPs are both minimal, since their cardinality is 8, as prescribed by 
Theorem~\ref{theo:minimalEMPforPPN}.				
				
As for the example on the right of Figure \ref{fig:Ex-PPNs}, there is a direct link from the source to the sink, and in addition there are two paths from the source to the sink that have one internal node. It follows that $K1=2$ and $K2=1$. Hence, by Theorem~\ref{theo:minimalEMPforPPN}, the cardinality of a minimal EMP is $6$. The minimal EMP, which is the only valid EMP,  is given by $ \mathcal{B} = \{1, 2, 3\}, \mathcal{C} = \{2, 3, 4\} $.

The next theorem, which will play a key role in our algorithm for the construction of a valid EMP,  shows under which simple conditions an edge can be identified if it is part of a PPN whose other paths are known.

\begin{theorem} \label{theo:knownpaths} 
				Let $G$ be a network matrix associated with a DAG. 
				Consider an edge $(i, j)$  and assume that there are other paths from $i$ to $j$. 
				If all edges of these other paths  are known,  
				then the transfer function $G_{ji}$ is generically identifiable provided node $i$ is excited and node $j$ is measured. 
\end{theorem}
\begin{proof}
Assume that the DAG is topologically sorted. 
				From Lemma III.1 in \cite{mapurunga-gevers-baza-CDC2022} we have: 
				\begin{align}
								G_{j i} = T_{j i} - 
								\sum_{k =i+1}^{j-1}  G_{j k} T_{k i}
								\label{eq:Tjiknownpaths}
				\end{align}
				or, alternatively, 
				\begin{align}
								G_{j i} = T_{j i} - 
								\sum_{k = i+1}^{j-1} T_{j k} G_{k i}								  			\label{eq:Tjiknownpaths2}
				.\end{align}
 If all edges from the other paths are known, 
				 this means that in (\ref{eq:Tjiknownpaths}) 
				 all $G_{j k}$ are known, and that the $T_{k i }$ can be reconstructed from these edges.		
				Since $j$ is measured, and $i$ is excited, $T_{j i}$ is also known. 
				Therefore, one can recover $G_{ji}$ from (\ref{eq:Tjiknownpaths}).  The same argument holds for  (\ref{eq:Tjiknownpaths2}).
\end{proof}

{We now introduce the important new concept of Induced Tree. We show that any node in a network is the source of a uniquely defined tree, which we define as the \textbf{induced tree (IT)}  of that node. ITs, together with PPNs,  will  be the two   building blocks of our decomposition algorithm for DAGs.}
\begin{definition}
				Consider a digraph $\mathcal{G}(\mathcal{V}, \mathcal{E})$ and let $v \in \mathcal{V}$ be any node of interest. {The digraph $\mathfrak{T}_v(\mathcal{V}_v, \mathcal{E}_v)$,} called the \textbf{induced tree (IT)} of node $v$, is formed as follows:
				the set $\mathcal{V}_v$ is composed of $v$ and all nodes of $\mathcal{V}$ that can be reached  from $v$ by exactly one directed path within $\mathcal{G}$,
				while the set $\mathcal{E}_v$ is formed by the edges in these unique paths that connect $v$ to $\mathcal{V}_v\setminus\{v\} $.								
				\label{def:induced-tree}
\end{definition}
{The 6-node tutorial example of Figure~\ref{fig:EduExample} in Section~\ref{sec:tutorial} illustrates Definition~\ref{def:induced-tree} since it lists the ITs of all 6 nodes in Table~\ref{tab:EduExample}.
}

Observe that an induced tree, as just defined, has the structure and the properties of a directed tree, since removal of any edge of the induced tree makes it disconnected. Thus, it is generically identifiable under the conditions of Theorem~\ref{theo:iso-trees}, with an EMP {of minimal cardinality, namely $n$,} the number of nodes.
The reason for introducing this concept will become apparent in the next section, where we show that any DAG can be decomposed into a set of induced trees and parallel paths networks, a property that would not hold if one just used trees instead of induced trees. {Note that, for brevity of notation, we shall in future  often just use the notation $\mathfrak{T}_v$ to denote the IT of node $i$, since the sets $\mathcal{V}_v$ and $\mathcal{E}_v$ are unambiguously defined by the IT of node $v$.}


\begin{lemma}\label{lem:ITinstep1}
Let $\mathcal{G}(\mathcal{V}, \mathcal{E})$ be a DAG. Consider an edge $(i,j)$ such that there are multiple paths from $i$ to $j$, and let $\mathfrak{T}_i(\mathcal{V}_i, \mathcal{E}_i)$ be the induced tree of node $i$. Then there exists a path $(i, k_i, k_{i+1}, \ldots, j)$ such that $(i,k_i) \in \mathfrak{T}_i $.
\end{lemma}
 \begin{proof}
Let  $k_i \in \mathcal{N}_i^+$ be on  a path to $j$ and suppose that $(i,k_i) \notin \mathfrak{T}_i$. This means that there is at least one other path from $i$ to $k_i$, which is therefore also on a path from $i$ to $j$. Let $l_i$ be the first node on that path; hence $l_i \in \mathcal{N}_i^+$. Then we must have $(i,l_i) \in  \mathfrak{T}_i$. Otherwise, we could repeat the same argument  as above for the edge $(i,l_i)$, each such step generating another element in $\mathcal{N}_i^+$. Since $\mathcal{N}_i^+$ is finite, there exists at least one node in $\mathcal{N}_i^+$ which is in $ \mathfrak{T}_i$ and which is on a path to $j$.
\end{proof}

Building on Lemma~\ref{lem:ITinstep1}, we now prove the following theorem, which will be the cornerstone of our decomposition algorithm for DAGs.

\begin{theorem}\label{theo:ITinmultipaths}
Let $\mathcal{G}(\mathcal{V}, \mathcal{E})$ be a DAG. Consider an edge $(i,j)$ such that there are multiple paths from $i$ to $j$. Then there exists a path $(i, k_i, k_{i+1}, \ldots, j)$ such that all edges in this path belong to induced trees.
\end{theorem}
 \begin{proof}
 By Lemma~\ref{lem:ITinstep1} there exists a node $k_i \in \mathcal{N}_i^+$ such that $k_i$ has a path to node $j$ and $(i,k_i) \in 
 \mathfrak{T}_i$. Now consider $\mathcal{N}_{k_i}^+$. There exists at least one out-neighbor $k_{i+1} \in \mathcal{N}_{k_i}^+$ which has a path to $j$. If the edge $(k_i,k_{i+1})$ is the only path from $k_i$ to $k_{i+1}$, then $(k_i,k_{i+1}) \in \mathfrak{T}_{k_i}$. If there are other paths from  $k_i$ to $k_{i+1}$, then by Lemma~\ref{lem:ITinstep1} there exists another node $l_{i+1} \in \mathcal{N}_{k_i}^+$ such that $l_{i+1}$ lies on a path to $j$ and $(k_i,l_{i+1}) \in \mathfrak{T}_{k_i}$. The result follows by repeating this argument until node $j$ is reached.
\end{proof}

In this section, we have introduced  two network topologies: ITs and PPNs. We have  given necessary and sufficient generic identifiability conditions for ITs and PPNs,  as well as the cardinality of a corresponding minimal EMP. {We have also developed a number of properties of ITs and PPNs that will play a key role in the decomposition of a DAG into a set of smaller subnetworks with these two specific structures.}

%

\section{Decomposing a DAG into induced trees and parallel paths networks}
\label{sec:decDAGs}

In this section we  show how a DAG can be decomposed into a set of subdigraphs that have the structure of an induced tree or a PPN, for which it is known how to construct valid and even minimal  EMPs,  as shown in Section~\ref{sec:structures}. We  then show under what conditions these EMPs, that are valid  for these subdigraphs, can, together, constitute a valid EMP for the whole digraph. To start, we formally define a subdigraph.
\begin{definition}
				Consider a digraph $\mathcal{G}(\mathcal{V}, \mathcal{E})$.
				A digraph $\mathcal{G}_A(\mathcal{V}_A, \mathcal{E}_A)$ is said to be a subdigraph of $\mathcal{G}$ if
				$\mathcal{V}_A \subseteq \mathcal{V}$ and
				$\mathcal{E}_A \subseteq \mathcal{E}$. 
				\label{def:subdigraph}
\end{definition}
We now present the notion of a covering set of a digraph by a set of subdigraphs.

\begin{definition}
				Consider a digraph $\mathcal{G}(\mathcal{V}, \mathcal{E})$ and let $\mathcal{G}_j(\mathcal{V}_j, \mathcal{E}_j)$ be a subdigraph of $\mathcal{G}$ indexed by $j$.  
				A set of subdigraphs $\{\mathcal{G}_j(\mathcal{V}_j, \mathcal{E}_j)\}_{j = 1}^k $ is said to be a \textbf{complete cover} of a digraph $\mathcal{G}(\mathcal{V}, \mathcal{E})$ 
				if $\mathcal{V} = \cup_{j = 1}^k \mathcal{V}_j$ 
				and $\mathcal{E} = \cup_{j = 1}^k \mathcal{E}_j$. 
				\label{def:complete-cover}
\end{definition}

Our first main result is to show that any DAG can be decomposed into a set of subdigraphs with topology of induced trees and parallel paths networks. 
\begin{theorem}
				Any directed acyclic graph (DAG) $\mathcal{G}\left(\mathcal{V}, \mathcal{E} \right)$ can be completely covered by a set of induced trees and parallel paths networks. 
				\label{theo:DAG-covered-by-trees-PPN}
\end{theorem}
\begin{proof}
								{For every node $v \in \mathcal{V}$, consider its corresponding induced tree $\mathfrak{T}_v(\mathcal{V}_v, \mathcal{E}_v)$. Each one of these ITs is a directed tree because it has a root, namely node $v$ 
								\cite[page 29]{RuohonenGraphTheory2013}. }
				{Let $\mathcal{G}_T(\mathcal{V}_T, \mathcal{E}_T)$ be the subdigraph of $\mathcal{G}\left(\mathcal{V}, \mathcal{E} \right)$ formed by all these induced trees for all $v \in \mathcal{V}$.}				
				Observe that it may not cover all of $\mathcal{G}\left(\mathcal{V}, \mathcal{E} \right)$. 
				
				{Thus, let  $\overline{\mathcal{E}} := \mathcal{E} \setminus \mathcal{E}_T$ be the set of edges that are not covered by $\mathcal{G}_T(\mathcal{V}_T, \mathcal{E}_T)$		and consider any edge $(i, j) \in \overline{\mathcal{E}}$. }
Since $(i,j) \in \overline{\mathcal{E}}$, it follows that there is more than one path from $i$ to $j$; otherwise node $j$ would be in an induced tree originating from node $i$. Take  two  paths from $i$ to $j$, one being the edge  $(i,j)$. 
				{Together these two paths form a parallel paths network, which we denote $\mathcal{PP}_{ij}$. 
				Repeat the same procedure for every edge in $\overline{\mathcal{E}}$ and denote by $\mathfrak{P} := \{\mathcal{PP}_{ij} | (i,j) \in \overline{\mathcal{E}}\}$ the set of all these parallel paths networks. 	The combination of $\mathcal{G}_T(\mathcal{V}_T, \mathcal{E}_T)$ and $\mathfrak{P}$ yields a 
 complete cover of $\mathcal{G}\left(\mathcal{V}, \mathcal{E} \right)$.} Note that, by the procedure adopted here, our algorithm only deals  with 2-paths PPNs.
\end{proof}
Any DAG  can thus be decomposed into a set of induced trees and parallel paths networks that jointly cover the whole DAG, but this covering is by no means unique. 

In  Section~\ref{sec:structures} we have shown that valid (and even minimal) EMPs are easy to construct for trees and PPNs.  However, these necessary and sufficient conditions, and the EMPs that are based on them, are valid for these structures taken in isolation. 
When these same structures are part of a more  complex network, these EMPs may no longer be valid. To address this question, we 
 extend Definition~\ref{def:EMP} to deal with EMPs that are valid for subdigraphs. 

\begin{definition}
Consider a digraph $\mathcal{G}(\mathcal{V}, \mathcal{E})$ and a subdigraph $\mathcal{G}_A(\mathcal{V}_A, \mathcal{E}_A)$ of $\mathcal{G}(\mathcal{V}, \mathcal{E})$. Consider an EMP that is valid for the subdigraph $\mathcal{G}_A(\mathcal{V}_A, \mathcal{E}_A)$ when all vertices and edges
from $\mathcal{G}(\mathcal{V}, \mathcal{E})$ that do not belong, respectively,
to $\mathcal{V}_A$ and  $\mathcal{E}_A$ have been removed. This EMP is said to be valid for the isolated subdigraph 
$\mathcal{G}_A(\mathcal{V}_A, \mathcal{E}_A)$. If the same EMP is valid when $\mathcal{G}_A(\mathcal{V}_A, \mathcal{E}_A)$ is embedded into $\mathcal{G}(\mathcal{V}, \mathcal{E})$, then this EMP is said to be valid for the embedded $\mathcal{G}_A(\mathcal{V}_A, \mathcal{E}_A)$.
\label{def:EMPsubemb}
\end{definition}

\begin{figure}[h!]
				\centering
				\includegraphics[width=0.5\columnwidth]{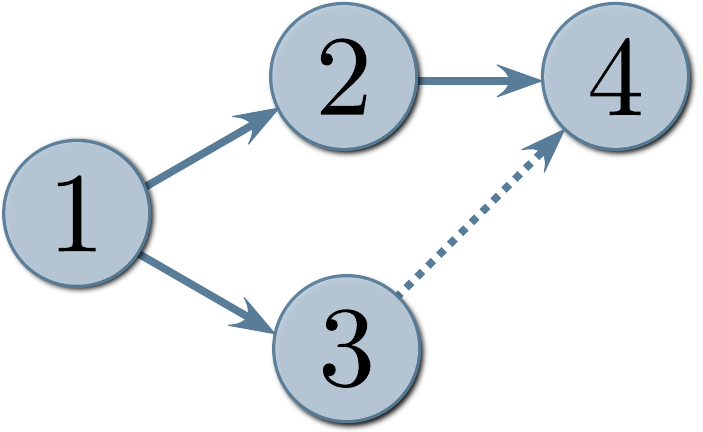}
				\caption{A DAG with an embedded tree in it. The tree is formed by nodes $1, 2, 3, 4$ and edges $(1, 2), (1, 3), (2, 4)$.}
				\label{fig:tree-example}
\end{figure}

Designing a valid EMP for an embedded digraph is usually a harder task than obtaining a valid EMP for its isolated counterpart. 
We illustrate this with the following example. 
Consider the DAG depicted in Figure \ref{fig:tree-example}.

Consider first that the dotted edge $(3,4)$  has been removed from the DAG, and that one wants to identify the  tree: $\mathcal{G}_T(\mathcal{V}_T, \mathcal{E}_T)$, with $\mathcal{V}_T = \{1, 2, 3, 4\}$ and $\mathcal{E}_T = \{(1, 2), (1, 3), (2, 4)\}$. 
From Theorem \ref{theo:iso-trees}, we know that a valid EMP for this tree would be
\emp{1}: 
$\mathcal{B}_1 = \{1, 2\}$,  
$\mathcal{C}_1 = \{3, 4\}$.
However, when  the edge $(3,4)$ is present, so that the tree is embedded in the DAG, this EMP is no longer valid for the identification of the tree - only the edges $(1,3)$ and $(2,4)$ can be identified in this case.

As this example illustrates, a valid EMP for a digraph taken in  isolation may no longer be valid when the digraph is embedded into a larger network. In Theorem~\ref{theo:DAG-covered-by-trees-PPN} we have shown that a DAG can be completely covered by ITs and PPNs, and in Section~\ref{sec:structures} we have shown how to construct valid (and actually minimal) EMPs for those structures taken in isolation. It is therefore important to know under which conditions these EMPs, that are valid for these subdigraphs, remain valid when they are embedded in the larger DAG. Sufficient conditions for this have been given in \cite{mapurunga-gevers-bazanella-CDC2024}. Here we recall the main result of \cite{mapurunga-gevers-bazanella-CDC2024} that addresses this problem.

\begin{theorem}\cite{mapurunga-gevers-bazanella-CDC2024} \label{theo:no-path-decomp}
				Consider a  digraph $\mathcal{G}(\mathcal{V}, \mathcal{E})$ and its associated
				network matrix $G$.
				Let $\mathcal{V}_A$ be a subset of connected nodes of $\mathcal{V}$,
				and ${\mathcal{V}_B} \triangleq \mathcal{V}\setminus \mathcal{V}_A$, 
				Let $\mathcal{G}_A(\mathcal{V}_A, \mathcal{E}_A)$ be the subdigraph of $\mathcal{G}$ defined by the subset of nodes $\mathcal{V}_A$ and 
				all edges $\mathcal{E}_A$ that link them, 
				and let $G_A$ be the associated subnetwork matrix.
				Assume that an EMP that is valid for the isolated subdigraph $\mathcal{G}_A$ has been obtained.
				This same EMP is also valid for the embedded subdigraph $\mathcal{G}_A$
				if  at least one of the following conditions holds.\\
				(1) There is no path starting in $\mathcal{V}_A$, passing through $\mathcal{V}_B$, and returning to $\mathcal{V}_A$. \label{theo:item:zero-path} \\
				(2) All paths that leave $\mathcal{V}_A$ and return to $\mathcal{V}_A$ are known.\label{theo:item:known-path}
\end{theorem}								
Observe that this theorem  applies to any network (not just DAGs) as it does not impose any topological structure on the associated graph. 
This result gives a sufficient condition under which we can determine if all transfer functions embedded in a subnetwork can be identified by applying an EMP that is valid for that subnetwork taken in isolation.

The following {new result}  shows that any induced tree embedded in a general network can always be identified if we apply to it an EMP that is valid for this tree taken in isolation, because condition 1) of Theorem~\ref{theo:no-path-decomp} always applies.
\begin{corollary}
				Consider a  digraph $\mathcal{G}(\mathcal{V}, \mathcal{E})$ and its associated
				network matrix $G$.
								Let $v \in \mathcal{V}$ be a node of interest and 
								let $\mathfrak{T}_v \left(\mathcal{V}_v, \mathcal{E}_v  \right) $ be its associated induced tree. 			
				If an EMP is valid for the identification of $\mathfrak{T}_v$ taken in isolation, it remains valid for the identification of $\mathfrak{T}_v$ when it is embedded in the digraph $\mathcal{G}(\mathcal{V}, \mathcal{E})$.
				\label{theo:decomptree}
\end{corollary}
\begin{proof}
An IT is defined as having at most one path between any two nodes in $\mathcal{V}_v$; see Definition \ref{def:induced-tree}. As a result, in an IT no path can exit the tree and return to it. Hence condition (1) of Theorem~\ref{theo:no-path-decomp} is always satisfied.		
\end{proof}			        			        

This result shows that for identifying the transfer functions of an induced tree embedded in a more general network it is sufficient to apply a valid EMP to it, i.e. one described in Theorem \ref{theo:iso-trees}. 
Notice that, just like Theorem~\ref{theo:no-path-decomp}, this 
 result holds for induced trees embedded in any network, regardless of its topology. 

Finally, the next Corollary follows immediately from item 2 of Theorem \ref{theo:no-path-decomp}.
\begin{corollary}
Consider a  digraph $\mathcal{G}(\mathcal{V}, \mathcal{E})$ and its associated
				network matrix $G$.	Let $\mathfrak{P}$ be a parallel paths network embedded in $\mathcal{G}$. 
				If a valid EMP is applied to $\mathfrak{P}$ and all paths that leave from $\mathfrak{P}$ and return to it are known,
				then the associated transfer functions from $\mathfrak{P}$ are generically identifiable. 
				\label{theo:parallel-path-decomp}
\end{corollary}
We have shown in Theorem~\ref{theo:DAG-covered-by-trees-PPN} that a DAG can be completely covered by a set of digraphs composed of ITs and PPNs. 
For ITs, Corollary~\ref{theo:decomptree} has shown that 
a valid EMP for an IT remains valid  when it is embedded into a complex network.
As for PPNs, Corollary~\ref{theo:parallel-path-decomp} has shown that 
one needs to guarantee that every path that leaves the PPN and returns to it is known. 
The results of Corollary~\ref{theo:decomptree} and Corollary~\ref{theo:parallel-path-decomp} are 
 instrumental for the derivation of an algorithm that produces an EMP with which all modules embedded in a DAG can be identified. This algorithm is developed in the next section.


%

\section{Design algorithm of a valid EMP for a DAG}
\label{sec:decalg}


In this section we propose an algorithm for the design of valid EMPs for DAGs by first decomposing  the DAG into subsets of  induced trees (ITs) and parallel paths networks (PPNs) which, together, completely cover the DAG. We then apply valid EMPs, defined locally, to these ITs and PPNs in such a way that these local EMPs remain valid when these local ITs and PPNs are embedded in the complete DAG. 

We first describe the Algorithm, and we then present two theorems that establish that the EMP constructed by the Algorithm is valid for the considered EMP.

\begin{algorithm}[H]
				\caption{Synthesis of a valid EMP for  a DAG}
				\label{alg:vEMPDAG}
				\textbf{Input}: Network with DAG topology $\mathcal{G}_0\left( \mathcal{V}_0, \mathcal{E}_0 \right)$ \\
				\textbf{Output}: EMP $\left(\mathcal{B},\mathcal{C}\right)$ valid for the whole network
				\begin{algorithmic}[1]
								\State Initialize an empty graph $\mathcal{G}(\mathcal{V},\mathcal{E})$
								and the empty sets $\mathcal{B}$, $\mathcal{C}$
							 \For{$i \in \mathcal{V}_0$}
							 \State $\mathfrak{T}_i (\mathcal{V}_i, \mathcal{E}_i) \gets $ Find the induced tree of node $i$;
							 \State $\mathcal{G} (\mathcal{V}, \mathcal{E})\gets  \mathcal{G} (\mathcal{V}, \mathcal{E}) \cup \mathfrak{T}_i(\mathcal{V}_i, \mathcal{E}_i)$
							\EndFor
							\State Denote $ {\mathcal G}_T({\mathcal V}_T, {\mathcal E}_T) = \mathcal{G} (\mathcal{V}, \mathcal{E})$
							\State $\mathcal{K} \gets $ The indices of the least number of induced trees such that 
							$\cup_i \mathfrak{T}_i  (\mathcal{V}_i, \mathcal{E}_i)= \mathcal{G}_{\mathcal{T}}(\mathcal{V}_{\mathcal{T}},\mathcal{E}_{\mathcal{T}})$) 
							\For {$i \in  \mathcal{K}$}
							   \State $\mathcal{B} \gets i$ 
								 \State $\mathcal{C} \gets$ the sinks of $\mathfrak{T}_i$ 
								 \State $\mathcal{B} \gets$ all internal nodes $k$ from $\mathfrak{T}_i$ such that there is an edge leaving $\mathcal{G}_T(\mathcal{V}_T, \mathcal{E}_T)$ at node $k$
								\State $\mathcal{C} \gets $ all internal nodes $k$ from $\mathfrak{T}_i$ such that there is an incoming edge coming from outside $\mathcal{G}_T(\mathcal{V}_T, \mathcal{E}_T)$ at node $k$								 \State Complement $\mathcal{B}$ and $\mathcal{C}$ so that the conditions of Theorem~\ref{theo:iso-trees} are satisfied for the identification of $\mathfrak{T}_i$
							\EndFor
							\State Define EMP$_{DAG}$ by $(\mathcal{B}, \mathcal{C})$
										\end{algorithmic}
\end{algorithm}
\ni{\bf Comment:} The ITs $\mathfrak{T}_i (\mathcal{V}_i, \mathcal{E}_i)$ defined for each node in step 3 often cover one another, as our examples will illustrate. Thus, in step 7, one should select the smallest possible number of ITs $\mathfrak{T}_i (\mathcal{V}_i, \mathcal{E}_i)$ that cover ${\mathcal G}_T({\mathcal V}_T, {\mathcal E}_T)$.
The EMP defined by  $(\mathcal{B}, \mathcal{C})$ in step 15 of the Algorithm has been designed solely on the basis of the ITs that compose ${\mathcal G}_T({\mathcal V}_T, {\mathcal E}_T)$.  However, for every one of these ITs, the Algorithm has not just selected any EMP that satisfies the conditions of Theorem~\ref{theo:iso-trees}, but one that obeys steps 11 and 12. {This is a way of keeping the cardinality of the resulting EMP low, as we will illustrate in the tutorial example of section~\ref{sec:tutorial}.}

We now show that, by this particular selection, the EMP$_{DAG}$ defined in step 15 also allows the identification of the PPNs that result from the decomposition provided by Theorem~\ref{theo:DAG-covered-by-trees-PPN}. Referring to the proof of that Theorem, we need to show that all edges $(i, j) \in \overline{\mathcal{E}}$, which support a PPN, are also identifiable with the EMP$_{DAG}$       defined for the ITs in the particular way described by the Algorithm. To prove this, we first establish the following result.

\begin{theorem}\label{result1}
Consider a DAG $\mathcal G(\mathcal V, \mathcal E)$, its IT cover ${\mathcal G}_T({\mathcal V}_T, {\mathcal E}_T)$, and an  $EMP_{IT}$  designed such that all edges in ${\mathcal E}_T$ have been identified. Consider any edge $(j,i) \in \bar {\mathcal E}:= {\mathcal E} \setminus {\mathcal E}_T$, and assume that there are multiple paths from node $j$ to node $i$. 
Let ${\mathcal G}_{ij}({\mathcal V}_{ij}, {\mathcal E}_{ij})$ be the digraph formed by all these paths from $j$ to $i$, ${\mathcal P}_{ij}$ the set of these paths, 
and let $\bar {\mathcal E}_{ij}$ be the set of all unknown edges in ${\mathcal E}_{ij}$. Then there exists an edge $(k,m) \in \bar {\mathcal E}_{ij}$ that belongs to a  PPN whose other edges  are all known, because they belong to ${\mathcal G}_T({\mathcal V}_T, {\mathcal E}_T)$.
Therefore, this edge $(k,m)$ is identifiable provided node $k$ is excited and node $m$ is measured.
\end{theorem}
\ni {\bf Proof:} Consider an edge $(k,m) \in \bar {\mathcal E}_{ij}$.
Denote by ${\mathcal P}_{km}$ the set of all paths from node $k$ to node $m$ in $\mathcal G(\mathcal V, \mathcal E)$, and by $\kappa({\mathcal P}_{km})$ the number of unknown edges in  ${\mathcal P}_{km}$, i.e. the number of edges in ${\mathcal P}_{km}$ that belong to $\bar {\mathcal E}_{ij}$.
By Theorem~\ref{theo:ITinmultipaths}, we know that, for every edge $(k,m) \in \bar {\mathcal E}_{ij}$, there exists at least one path $(k,k+1,\ldots,m)$ whose edges are all known because they belong to ${\mathcal G}_T({\mathcal V}_T, {\mathcal E}_T)$. Two situations can arise.\\
Case 1: All paths in ${\mathcal P}_{km}$, except for the direct edge $(k,m)$, have this property, i.e. their edges are all known because they belong to ${\mathcal G}_T({\mathcal V}_T, {\mathcal E}_T)$. In such case, identifiability of the edge $(k,m)$  follows from Theorem~\ref{theo:knownpaths}, provided node $i$ is excited and node $j$ is measured.\\
Case 2: There exists a path in ${\mathcal P}_{km}$, other than the edge $(k,m)$,  that contains an unknown edge, say $(l,n)$. Now consider the set ${\mathcal P}_{ln}$ of all paths from node $l$ to node  $n$. Applying Theorem~\ref{theo:ITinmultipaths} again, we know that there  exists at least one path $(l,l+1,\ldots,n)$ whose edges are all known. If the edges in the other paths from $l$ to $n$ are not all known, we repeat the same argument to another unknown edge in ${\mathcal P}_{ln}$. However, observe that  ${\mathcal P}_{ln}$ is a subset of ${\mathcal P}_{km}$, which is a finite set, and therefore $\kappa({\mathcal P}_{ln}) < \kappa({\mathcal P}_{km})$. As a result, the number of possible unknown edges keeps shrinking. 
Repeating the same argument again will lead to an edge, say  $(u,v)$, with $\kappa({\mathcal P}_{uv})=1$, 
where the edges in the only other path from $u$ to $v$ are known because they belong to ${\mathcal G}_T({\mathcal V}_T, {\mathcal E}_T)$.
Thus the edge $(u,v)$ can be identified by Theorem~\ref{theo:knownpaths} provided $u$ is excited and $v$ is measured, which proves the result.
\cqfd

 The proof that EMP$_{DAG}$ defined in step 15 of the Algorithm defines a valid EMP for the DAG now follows from the following main result.
 
 \begin{theorem}\label{algotheorem}
Consider a DAG $\mathcal G(\mathcal V, \mathcal E)$, its IT cover ${\mathcal G}_T({\mathcal V}_T, {\mathcal E}_T)$, and the  EMP$_{DAG}$  designed by the Algorithm (see step 15). Let $ \bar {\mathcal E}:= {\mathcal E} \setminus {\mathcal E}_T$ be the set of all edges in $\mathcal G(\mathcal V, \mathcal E)$ that are not in ${\mathcal E}_T$. Then all edges 
$(i,j) \in \bar {\mathcal E}$ can be identified by applying this EMP$_{DAG}$ to the DAG.
\end{theorem}
\ni {\bf Proof:} The proof consists of two parts.\\
$\bullet$ {\bf Part 1.} We first show that all edges in $ \bar {\mathcal E}$ can be identified using the procedure described in the proof of Theorem~\ref{result1}. \\
Consider any edge $(i,j) \in  \bar {\mathcal E}$. If there is only one other path from node $i$ to node $j$, then all edges of this other path are known because, by Theorem~\ref{theo:ITinmultipaths}, 
they belong to ${\mathcal E}_T$. 
Therefore, by Theorem~\ref{theo:knownpaths},  
$(i,j)$ is identifiable provided node $i$ is excited and node $j$  is measured. \\
Assume now that  there are multiple paths from node $i$ to node $j$. Denote by ${\mathcal P}_{ij}$ the set of all these paths from node $i$ to node $j$ in $\mathcal G(\mathcal V, \mathcal E)$, and by $\kappa({\mathcal P}_{ij})$ the number of unknown edges in  ${\mathcal P}_{ij}$, i.e. the number of edges in ${\mathcal P}_{ij}$ that belong to $\bar {\mathcal E}_{ij}$. 
By Theorem~\ref{result1}, there exists an edge in ${\mathcal P}_{ij}$, say  $(k,m)$,  that can be identified provided node $k$ is excited and node $m$ is measured, in addition to EMP$_{DAG}$ being applied. After identification of $(k,m)$, $\kappa({\mathcal P}_{ij})$ is reduced by 1. Repeating the same procedure to any other  edge, say $(u,v)$,  in ${\mathcal P}_{ij}$ with node $u$ being excited and node $v$ measured produces another known edge in ${\mathcal P}_{ij}$, thus again reducing $\kappa({\mathcal P}_{ij})$  by 1. This procedure can be repeated until $\kappa({\mathcal P}_{ij}) = 0$.
The same procedure can be applied to all unknown edges in $ \bar {\mathcal E}$, which proves Part 1.\\
$\bullet$ {\bf Part 2.} We have shown in Theorem~\ref{result1} that the identification of each edge $(k,m) \in \bar {\mathcal E}_{ij}$ is based on its inclusion in a PPN in which all other paths from node $k$ to node $m$ belong to ${\mathcal E}_T$, and are hence known by the application of EMP$_{DAG}$. The edge $(k,m)$ is then identifiable, on the basis of Theorem~\ref{theo:knownpaths},
provided node $k$ is excited and node $m$ is measured. Now it follows from steps 11 and 12 of the Algorithm that node $k$ is already included in ${\mathcal B}$ and that node $m$ is already included in ${\mathcal C}$, which together form EMP$_{DAG}$. This proves the result. \cqfd

\ni{\bf Important comment:} The Algorithm produces an EMP$_{DAG}$ that is entirely based on the ITs contained in the DAG, i.e. it is a valid EMP for ${\mathcal G}_T({\mathcal V}_T, {\mathcal E}_T)$. However, as a result of the particular choices made in steps 11 and 12, this EMP turns out to be also valid for all edges of $ \bar {\mathcal E}$ which belong to PPNs. This is proved by Theorem~\ref{result1} and Theorem~\ref{algotheorem}. These proofs crucially depend on the decomposition of the DAG into ITs and PPNs, that results from Theorem~\ref{theo:DAG-covered-by-trees-PPN}, and on the identifiability conditions  for isolated and embedded PPNs that result from Theorem~\ref{theo:generalparallel} and Corollary~\ref{theo:parallel-path-decomp}, respectively.

\section{A tutorial example}\label{sec:tutorial}

\begin{figure}[h!]
				\centering
				\includegraphics[width=0.5\columnwidth]{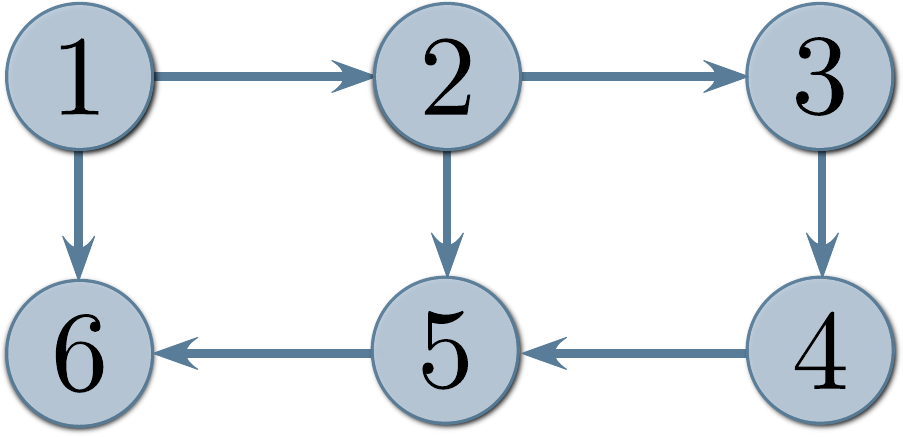}
				\caption{A simple 6-node  DAG.}
				\label{fig:EduExample}
\end{figure}

Here we present a simple tutorial example, whose purpose is to help the reader understand the details of each step of the algorithm. In the next section, we shall apply the algorithm to a moderate size DAG, without detailing these steps.
Consider the network of Figure~\ref{fig:EduExample}. It has the structure of a DAG, with 6 nodes and 7 edges.

We list the induced trees in Table~\ref{tab:EduExample}.
\footnotesize{
\begin{table}[h!]
				\caption{Induced trees for the digraph of Figure \ref{fig:EduExample}. } 
				\label{tab:EduExample}
\scriptsize{					\begin{tabular}{|c |c| c| c | c | c |}
								\hline
                \thead{Tree} & \thead{{$\mathcal{V}_{i}$}} & \thead{{$\mathcal{E}_{i}$}} & Tree & \thead{{$\mathcal{V}_{i}$}} & \thead{{$\mathcal{E}_{i}$}}\\\hline
                $\mathfrak{T}_1$ & $1, 2, 3, 4$ & $(1, 2), (2, 3), (3, 4) $ &
                $\mathfrak{T}_2$ & $2, 3, 4 $ & $(2, 3), (3, 4) $ \\\hline
								$\mathfrak{T}_3$ &
                \makecell{$3, 4, 5, 6$\\} & 
                \makecell{$(3, 4), (4, 5), (5, 6)$} &
								$\mathfrak{T}_4$ & 
								\makecell{$4, 5, 6$} 
								& \makecell{$(4, 5), (5, 6)$} \\\hline
								$\mathfrak{T}_5$ & \makecell{$5, 6$} & 
                \makecell{$(5, 6)$} & & &  \\\hline
								\end{tabular}
}				
\end{table}


It follows that $ \mathfrak{T}_1$ and  $\mathfrak{T}_3$ define the largest possible cover of $\mathcal{G}_0\left( \mathcal{V}_0, \mathcal{E}_0 \right)$, i.e. $ \mathcal{G}_T (\mathcal{V}_T,\mathcal{E}_T) = \mathfrak{T}_1  \cup \mathfrak{T}_3$. We observe that all edges of $\mathcal{E}_0$ are covered by 
$ \mathcal{G}_T (\mathcal{V}_T,\mathcal{E}_T)$ except for edges (1,6) and (2,5), which jointly define $\overline{\mathcal{E}}$.

We now define valid EMPs for these two ITs taken in isolation, following the rules defined in steps 9 to 12 of the Algorithm. This yields:\\
$\bullet$ for $ \mathfrak{T}_1$ : $\{1, 2\} \subset \mathcal{B}_1, \{4\} \subset  \mathcal{C}_1$, and $\{3\}$ in $\mathcal{B}_1$ or $\mathcal{C}_1$ \\
$\bullet$ for $ \mathfrak{T}_3$ : $\{3\} \subset \mathcal{B}_3 ,  \{5,6\} \subset \mathcal{C}_3$, and $\{4\}$ in $\mathcal{B}_3$ or $\mathcal{C}_3$ \\
Combining the requirements for the individual ITs yields an EMP 
defined by  ${\mathcal B} = \{1,2,3\}, {\mathcal C}= \{4,5,6\}$. This is a valid EMP for the identification of all edges in $ \mathfrak{T}_1$ and $ \mathfrak{T}_3$. Thus, all edges on the path (1,2,3,4,5,6) are identifiable with this EMP. Hence step 13 is not required here.

We now check that the edges $(1,6)$ and $(2,5)$ that belong to $\overline{\mathcal{E}}$ are also identifiable with this EMP. \\
$\bullet$ The edge (1,6) belongs to a 2-paths PPN formed by (1,6) and (1,2,3,4,5,6) where all edges on the path 
(1,2,3,4,5,6) are known.\\
$\bullet$ The edge (2,5) belongs to a 2-paths PPN formed by (2,5) and (2,3,4,5) where all edges on the path 
(2,3,4,5) are known. 

It follows from Theorem~\ref{theo:generalparallel} that the EMP defined above  is valid for these two PPNs, and that they thus allow the identification of the edges $(1,6)$ and $(2,5)$.
We conclude that the EMP, defined by  ${\mathcal B} = \{1,2,3\}$ and ${\mathcal C}= \{4,5,6\}$, is valid for the identification of the whole DAG. In addition, it is minimal since its  cardinality is 6, which is the number of nodes.

{\ni{\bf Comment on steps 11 and 12 of the Algorithm}\\
 In the definition of the EMPs for $ \mathfrak{T}_1$ and $ \mathfrak{T}_3$, we note that node $2$ has been put in $\mathcal{B}_1$ by rule 11, while node $5$ has been put in $\mathcal{C}_3$ by rule 12. If one only considers the identification of $ \mathfrak{T}_1$ and $ \mathfrak{T}_3$ without applying these two rules, 
  one could have put node $2$ in $\mathcal{C}_1$ and node $5$ in $\mathcal{B}_3$. This would yield the  EMP ${\mathcal B} = \{1,3,5\}, {\mathcal C}= \{2,4,6\}$, which is valid for the identification of $ \mathfrak{T}_1$ and $ \mathfrak{T}_3$. However, with this choice, the edge $(2,5)$ cannot be identified in the PPN $(2,3,4,5)$ using Theorem~\ref{theo:knownpaths}, and the same holds for the edge $(1,6)$. Thus, additional excitations or measurements would be required to identify these two paths, thus increasing the cardinality of the EMP. For example, adding node $2$ in ${\mathcal B} $ yields the valid EMP ${\mathcal B} = \{1,2,3,5\}, {\mathcal C}= \{2,4,6\}$; alternatively, adding node $5$ in 
 ${\mathcal C}$ yields the valid EMP ${\mathcal B} = \{1,3,5\}, {\mathcal C}= \{2,4,5,6\}$. Both have cadinality 7.}

\section{Example}
\label{sec:example}
 We now  apply the Algorithm  to the moderate size dynamical network depicted in Figure \ref{fig:dag1},  which has the structure of a DAG with 25 nodes. 

\begin{figure}[h!]
				\centering
				\includegraphics[width=0.7\columnwidth]{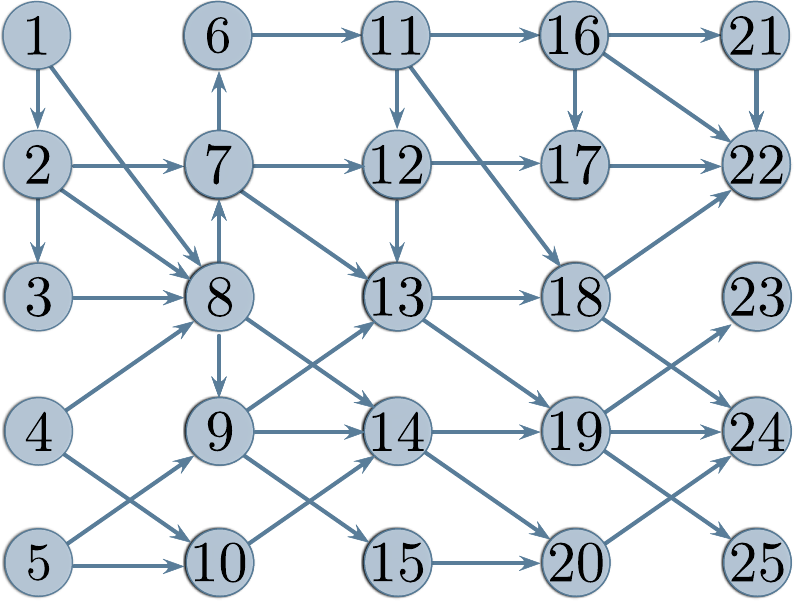}
				\caption{A 25-node DAG.}
				\label{fig:dag1}
\end{figure}

The first step in the algorithm is to construct an induced tree $\mathfrak{T}_i$ for every node $i$ in the network. 
The induced trees for all nodes of the digraph in Figure \ref{fig:dag1} are given in Table \ref{tab:itreesEx1}. Note that the sinks (nodes 22 to 25) have no induced trees.
\footnotesize{
\begin{table}[h!]
				\centering
				\caption{Induced trees for the digraph of Figure \ref{fig:dag1}. } 
				\label{tab:itreesEx1}
\scriptsize{					\begin{tabular}{|c |c| c|}
								\hline
			\thead{Tree} & \thead{$\mathcal{V}_{\mathfrak{T}_i}$} & \thead{$\mathcal{E}_{\mathfrak{T}_i}$} \\\hline
								$\mathfrak{T}_1$ & $1, 2, 3 $ & $(1, 2), (2, 3) $\\\hline
								$\mathfrak{T}_2$ & $2, 3 $ & $(2, 3) $ \\\hline
								$\mathfrak{T}_3$ &
								\makecell{$3, 6, 7, 8, 9, 11,$\\$ 15, 16, 21 $} & 
								\makecell{$(3, 8), (8, 7), (7, 6), (6, 11), (11, 16),$\\$(16, 21), (8, 9), (9, 15) $} \\\hline
								$\mathfrak{T}_4$ & 
								\makecell{$4, 6, 7, 8, 9, 10$\\$11, 15, 16, 21$} 
								& \makecell{$(4, 8), (4, 10), (8, 9), (8, 7), (7, 6), $\\$(6, 11), (11, 16), (16, 21),  (9, 15)$} \\\hline
								$\mathfrak{T}_5$ & \makecell{$5, 9, 10, 13, 15$\\$18, 22 $} & 
								\makecell{$(5, 9), (5, 10), (9, 13), (9, 15)$ \\ $(13, 18), (18, 22)$} \\\hline
								$\mathfrak{T}_6$ & 
								\makecell{$6, 11, 12, 13, 16 $ \\ $19, 21, 23, 25$} &
								\makecell{$(6, 11), (11, 12), (11, 16), (12, 13)$ \\ $(13, 19), (16, 21), (19, 23), (19, 25)$ } \\\hline
								$\mathfrak{T}_7$ & $6, 7, 11, 16, 21 $ & $(7, 6), (6, 11), (11, 16), (16, 21)$ \\\hline
								$\mathfrak{T}_8$ &
								\makecell{$6, 7, 8, 9, 11, 15 $\\$16, 21$} & 
								\makecell{$(8, 7), (8, 9), (7, 6), (9, 15), (6, 11)$ \\ $(11, 16), (16, 21)$ }\\\hline
								$\mathfrak{T}_9$ & \makecell{$9, 13, 14, 15 $ \\ $18, 22$} & 
								\makecell{$(9, 13), (9, 14), (9, 15), (13, 18)$ \\ $(18, 22)$} \\\hline
								$\mathfrak{T}_{10}$ & $10, 14, 19, 20, 23, 25 $ & 
								\makecell{$(10, 14), (14, 19), (14, 20), (19, 23)$\\$(19, 25)$} \\\hline
								$\mathfrak{T}_{11}$ & 
								\makecell{$11, 12, 13, 16, 19, 21$\\$23, 25$} &
								\makecell{$(11, 12), (11, 16), (12, 13), (13, 19), $\\
								$(16, 21), (19, 23), (19, 25)$} \\\hline
								$\mathfrak{T}_{12}$ & 
								\makecell{$12, 13, 17, 18, 19, 23 $\\ $25$} & 
								\makecell{$(12, 13), (12, 17), (13, 18), (13, 19), $\\$(19, 23), (19, 25)$} \\\hline
								$\mathfrak{T}_{13}$ & $13, 18, 19, 22, 23, 25 $ & 
								\makecell{$(13, 18), (13, 19), (18, 22), (19, 23),$\\$ (19, 25)$} \\\hline
								$\mathfrak{T}_{14}$ & $14, 19, 20, 23, 25 $ & $(14, 19), (14, 20), (19, 23), (19, 25)$ \\\hline
  							$\mathfrak{T}_{15}$ & $15, 20, 24 $ & $(15, 20), (20, 24)$ \\\hline
								$\mathfrak{T}_{16}$ & $ 16, 17, 21 $ & $(16, 17), (16, 21)$ \\\hline
								$\mathfrak{T}_{17}$ & $ 17, 22 $ & $(17, 22)$ \\\hline
								$\mathfrak{T}_{18}$ & $ 18, 22, 24 $ & $(18, 22), (18, 24)$ \\\hline
								$\mathfrak{T}_{19}$ & $ 19, 23, 24, 25 $ & $(19, 23), (19, 24), (19, 25)$ \\\hline
								$\mathfrak{T}_{20}$ & $ 20, 24 $ & $(20, 24)$ \\\hline
								$\mathfrak{T}_{21}$ & $ 21, 22 $ & $(21, 22)$ \\\hline
				\end{tabular}
}				
\end{table}

We update the initialized digraph $\mathcal{G}$ with the union of all induced trees for each node. This defines 
$ \mathcal{G}_T (\mathcal{V}_T,\mathcal{E}_T) = \cup ~\mathfrak{T}_i(\mathcal{V}_i, \mathcal{E}_i)$; see step 6 of the algorithm.
It is a subdigraph of $\mathcal{G}_0\left( \mathcal{V}_0, \mathcal{E}_0 \right)$, composed of all the green edges in Figure~\ref{fig:ExLargeAlgTrees}.

\begin{figure}[h!]
				\centering
				\includegraphics[width=0.7\columnwidth]{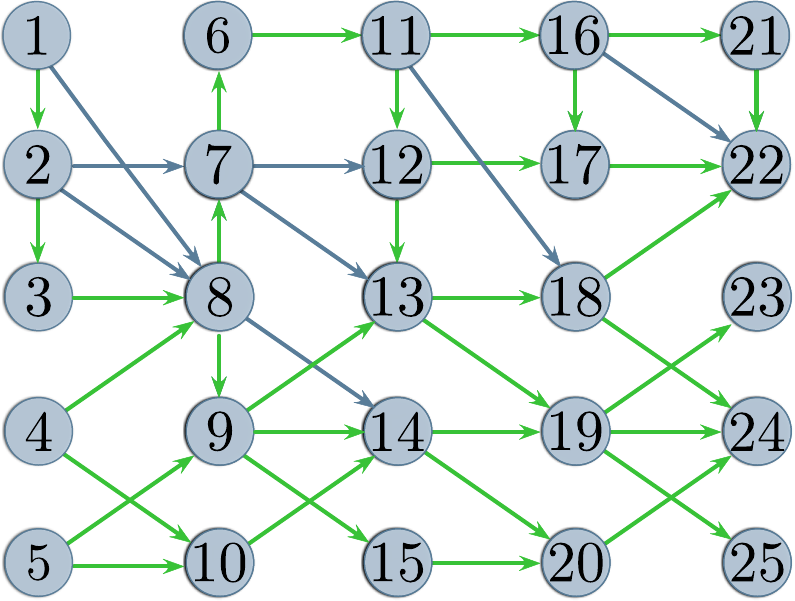}
				\caption{
								The edges depicted in green compose the digraph $\mathcal{G}_{\mathfrak{T}}(\mathcal{V}_{\mathfrak{T}},\mathcal{E}_{\mathfrak{T}})$  defined in step $6$ in the Algorithm. 								Each induced tree needs a specific EMP for identifiability of their edges. }
				\label{fig:ExLargeAlgTrees}
\end{figure}
We observe that this induced tree cover contains most of the edges of the DAG $\mathcal{G}_0\left( \mathcal{V}_0, \mathcal{E}_0 \right)$.
We then select the smallest possible number of induced trees capable of covering the subdigraph $ \mathcal{G}_T (\mathcal{V}_T,\mathcal{E}_T)$. To do this, we 
start by eliminating the induced trees that are contained into larger induced trees. 
For example, the induced tree of node 7 ($\mathfrak{T}_7$) is contained in the induced tree of node $8$ ($\mathfrak{T}_8$) given in Table~\ref{tab:itreesEx1}, i.e.~ all nodes and edges of $\mathfrak{T}_7$ are part of $\mathfrak{T}_8$. Thus $\mathfrak{T}_7$ can  be eliminated from the cover. 

With this elimination process, the set $ \mathcal{G}_T (\mathcal{V}_T,\mathcal{E}_T) = \cup~ \mathfrak{T}_i(\mathcal{V}_i, \mathcal{E}_i)$ can be covered by the  union of all directed trees induced by the nodes contained in $\mathcal{K} = \{1, 3, 4, 5, 6, 9, 10, 12, 15, 16, 17, 18, 19, 21\}$. It is again shown by the green edges in  Figure~\ref{fig:ExLargeAlgTrees} since this union of induced trees defined by $\mathcal K$ is a complete cover of $ \mathcal{G}_T (\mathcal{V}_T,\mathcal{E}_T)$.

The second part of the Algorithm is to design a valid EMP for these induced trees, following the procedure defined in steps 9 to 12 of the Algorithm.
Collecting the EMPs designed for each one of these ITs  yields the following EMP for $ \mathcal{G}_T (\mathcal{V}_T,\mathcal{E}_T) = \cup ~\mathfrak{T}_i(\mathcal{V}_i, \mathcal{E}_i)$:\\
\begin{itemize}
\item  ${\mathcal B} = \{1,2,3,4,5,6,7,8,9,10,11,12,15,16,17,18,19, 21\}$
\item $ {\mathcal C}= \{3,7,8,10,12,13,14,15,17,18,20,21,22,23,24,25\}$
\end{itemize}
The cardinality of this EMP is 34.

We now verify that all edges in $\overline{\mathcal{E}} =:  \mathcal{E}_{0} \setminus \mathcal{E}_{\mathfrak{T}}$, that are not covered by this induced tree cover, are identifiable by the above defined EMP.
It is easy to see on Figure~\ref{fig:ExLargeAlgTrees} that 
$\overline{\mathcal{E}} = \{(1,8), (2,7), (2,8), (7,12), (7,13), (8,14), (11,18), (16,22)\}$. In Table~\ref{tab:PPNsEn} we have listed these edges, together with the PPNs that they support. They are listed in the order in which they can be identified with the EMP defined above, based on the methodology described in the proof of Theorem~\ref{algotheorem}. 

\begin{table}[h!]
				\centering
				\caption{Parallel paths networks for each edge in $\overline{\mathcal{E}}$.}
				\label{tab:PPNsEn}
				\begin{tabular}{|c|c|c|c|}
								\hline
								\thead{Edge in $\overline{\mathcal{E}}$} &  The other paths \\\hline		
								$(2, 8)$   &  $(2,3,8)$\\\hline
								$(1, 8)$   & $(2,3,8)$ and $(1,2,3,8)$ \\\hline
								$(2, 7)$   &  $(2,8,7)$ and $(2,3,8,7)$ \\\hline
								$(7, 12)$  & $(7,6,11,12)$ \\\hline
								$(7, 13)$  & $(7,12,13)$ \\\hline
								$(8, 14)$  &  $(8, 9,14)$ \\\hline
								$(11, 18)$ & $(11,12,13,18)$ \\\hline
								$(16, 22)$ & $(16,17,22)$ and $(16,21,22)$ \\\hline
				\end{tabular}
\end{table}
We observe that each one of the edges $\overline{\mathcal{E}}$ belongs to a  PPN, whose other paths are known from the ITs by  application of the EMP defined above. Therefore, all edges in $\overline{\mathcal{E}}$ can be identified by application of Theorem~\ref{theo:knownpaths}.

Thus, the EMP for the induced trees, defined by the Algorithm,  is valid for the whole DAG. We repeat it here.
EMP$_{DAG}$ is defined by
\begin{itemize}
\item  ${\mathcal B} = \{1,2,3,4,5,6,7,8,9,10,11,12,15,16,17,18,19, 21\}$
\item $ {\mathcal C}= \{3,7,8,10,12,13,14,15,17,18,20,21,22,23,24,25\}$
\end{itemize}
We observe that the cardinality of this EMP for the whole DAG is 34, for a network that has 25 nodes. It can be shown that the cardinality of a valid EMP for this 25-node DAG is between 27 and 43.

{\textbf{Comments about the cardinality of the EMP}} \\
{Unless other objectives are being pursued, such as obtaining an EMP that minimizes some measure  of the estimated edges as in \cite{mapurunga-optimal-2021,mapurunga-bazanella-optimal-excitation-2023}, or obtaining one that minimizes the cost of the experiment as in \cite{JahandariMaterassi:2022,Bazanella&Gevers&Mapurunga:CDC2025}, one typically wants to achieve a valid EMP that is sparse, as measured by its cardinality. 
Lemma~\ref{cardinality} gives a lower bound, $n$, and an upper bound $2n-f-s$ for the cardinality of an EMP for any network, where $f$ is the number of sources and $s$ the number of sinks. While it is trivial, but uninteresting, to construct a valid EMP whose cardinality achieves the upper bound, there is at this point no way to compute a minimal EMP except by performing an exhaustive search which is unrealistic even for moderate size networks. In addition, there is no guarantee that there exists a minimal EMP whose cardinality achieves the lower bound of $n$.
In fact, in this particular example we know that the lower bound $n=25$ can not be achieved, because there 
are two nodes ($3$ and $21$) that must be both excited and measured, since they are 
dources and dinks at the same time - see Proposition 2.1. Hence the cardinality of a minimal EMP is at least $27$.}

{As a result we make no claim of minimality for  the EMP designed by Algorithm~\ref{alg:vEMPDAG} of Section~\ref{sec:decalg}, or that its cardinality will be close to the lower bound $n$. However, the algorithm contains  two features  that tend to reduce its cardinality, as explained at the end of the Introduction and in the Important comment at the end of Section~\ref{sec:decalg}, and as illustrated in the discussion of the tutorial example at the end of Section~\ref{sec:tutorial}.}

%
%

\section{Conclusion}
\label{sec:conclu}
We have presented an algorithm that synthesizes a valid EMP for Directed Acyclic Networks, with small cardinality. The algorithm is based on a novel decomposition procedure of the DAG into subdigraphs made up of trees and PPNs. Experimental conditions under which a tree can be identified have been known for some time. As for PPNs, we have given necessary and sufficient experimental conditions for their identification in our recent work \cite{mapurunga-gevers-bazanella-CDC2024}. In that work, we have also shown under which conditions an EMP that is valid for a subdigraph taken in isolation remains valid when this subdigraph is embedded in a larger digraph. These two results were essential for the decomposition method developed in the present work.

This decomposition has been made possible by the introduction of the new concept of Induced Tree, which has all the properties of a tree. We have shown that a DAG can be completely covered by subdigraphs that are either ITs or PPNs. Valid EMPs have been derived for these two structures, and conditions have been derived under which 
 these EMPs remain valid when the subdigraphs are put together. It remained to construct an algorithm that designs an EMP on the basis of this decomposition, that is valid for the whole DAG. Our algorithm is based on constructing an EMP for the set of induced trees, while taking into account the connections between these induced trees and the PPNs of the decomposition.  This is the main contribution of the present paper. 

With the results of the present paper, the design of a valid EMP for a DAG is now a very simple task. The procedure is almost entirely automatic, and our experience has shown that the cardinality of the EMPs that are produced by the algorithm is low, even though there is no claim of it being minimal.


\bibliographystyle{plain}
\bibliography{decbib, migrefs}

@book{RuohonenGraphTheory2013,
  title = {Graph Theory},
  author = {Ruohonen, Keijo},
  publisher = {Flooved.com},
  year = {2013},
  series = {Lecture Notes},
  urldate = {2023-04-12},
}

@ARTICLE{Materassi-Innocenti:2010,
  author={Materassi, Donatello and Innocenti, Giacomo},
  journal={IEEE Transactions on Automatic Control}, 
  title={Topological Identification in Networks of Dynamical Systems}, 
  year={2010},
  volume={55},
  number={8},
  pages={1860-1871},
  doi={10.1109/TAC.2010.2042347}}

@ARTICLE{Materassi-Salapaka:2012,
  author={Materassi, Donatello and Salapaka, Murti V.},
  journal={IEEE Transactions on Automatic Control}, 
  title={On the Problem of Reconstructing an Unknown Topology via Locality Properties of the Wiener Filter}, 
  year={2012},
  volume={57},
  number={7},
  pages={1765-1777},
  doi={10.1109/TAC.2012.2183170}}

@article{Pare-Beck-Basar:2020,
title = {Modeling, estimation, and analysis of epidemics over networks: An overview},
journal = {Annual Reviews in Control},
volume = {50},
pages = {345-360},
year = {2020},
issn = {1367-5788},
doi = {https://doi.org/10.1016/j.arcontrol.2020.09.003},
author = {Philip E. Par{\'e} and Carolyn L. Beck and Tamer Ba{\c s}ar}
}

@article{Munaro:2015,
	author = {Marques, Vinicius M. and Munaro, Celso J. and Shah, Sirish L.},
	doi = {10.1109/TASE.2015.2435897},
	journal = {IEEE Transactions on Automation Science and Engineering},
	number = {4},
	pages = {1525-1534},
	title = {Detection of Causal Relationships Based on Residual Analysis},
	volume = {12},
	year = {2015}}

@article{MattaSayed:2019,
	author = {Matta, Vincenzo and Sayed, Ali H.},
	doi = {10.1109/TIT.2018.2839192},
	journal = {IEEE Transactions on Information Theory},
	number = {1},
	pages = {622-646},
	title = {Consistent Tomography Under Partial Observations Over Adaptive Networks},
	volume = {65},
	year = {2019}}

@article{Ardakanianetall:2019,
	author = {Ardakanian, Omid and Wong, Vincent W. S. and Dobbe, Roel and Low, Steven H. and von Meier, Alexandra and Tomlin, Claire J. and Yuan, Ye},
	doi = {10.1109/TCNS.2019.2891002},
	journal = {IEEE Transactions on Control of Network Systems},
	number = {3},
	pages = {950-960},
	title = {On Identification of Distribution Grids},
	volume = {6},
	year = {2019}}

@article{NabaviChakrabortty:2016,
	author = {Seyedbehzad Nabavi and Aranya Chakrabortty},
	journal = {IEEE Transactions on Automatic Control},
	month = {February},
	number = {2},
	pages = {497-502},
	title = {A Graph-Theoretic Condition for Global Identifiability of Weighted Consensus Networks},
	volume = {61},
	year = {2016}}

@article{MaterassiSalapaka:2020,
	author = {Materassi, Donatello and Salapaka, Murti V.},
	journal = {IEEE Transactions on Automatic Control},
	number = {10},
	pages = {4138-4153},
	title = {Signal Selection for Estimation and Identification in Networks of Dynamic Systems: A Graphical Model Approach},
	volume = {65},
	year = {2020},
}

@article{mapurunga-bazanella-optimal-excitation-2023,
  title = {Optimal Excitation and Measurement Pattern for Cascade Networks},
  author = {Mapurunga, Eduardo and Bazanella, Alexandre Sanfelici},
  year = {2023},
  month = may,
  journal = {Automatica},
  volume = {151},
  pages = {110885},
  issn = {0005-1098},
  doi = {10.1016/j.automatica.2023.110885},
}

@article{JahandariMaterassi:2022,
	author = {Sina Jahandari and Donatello Materassi},
	journal = {IEEE Transactions on Automatic Control},
	month = {September},
	number = {9},
	pages = {4703-4716},
	title = {Optimal Selection of Observations for Identification of Multiple Modules in Dynamic Networks},
	volume = {67},
	year = {2022}}

@inproceedings{mapurunga-gevers-bazanella-CDC2024,
	author = {Mapurunga, Eduardo and Gevers, Michel and Bazanella, Alexandre S.},
	booktitle = { 63rd IEEE  Conference on Decision and Control, Milano},
	publisher = {IEEE},
  	month = dec,
	pages = {3147-3152},
	title = {Extending identifiability results from isolated networks to embedded networks},
	year = {2024},
}

@inproceedings{Bazanella&Gevers&Mapurunga:CDC2025,
	author = {Bazanella, Alexandre S. and Gevers, Michel and  Mapurunga, Eduardo.},
	booktitle = { 64th IEEE  Conference on Decision and Control, Rio de Janeiro},
	publisher = {IEEE},
  	month = dec,
	pages = {1364-1369},
	title = {Identifying dynamical networks with minimal cost},
	year = {2025},
}

@inproceedings{mapurunga-gevers-baza-CDC2022,
	author = {Mapurunga, Eduardo and Gevers, Michel and Bazanella, Alexandre S.},
	booktitle = {2022 {{IEEE}} 61st {{Conference}} on {{Decision}} and {{Control}} ({{CDC}})},
	doi = {10.1109/CDC51059.2022.9993388},
	issn = {2576-2370},
	month = dec,
	pages = {1616--1621},
	title = {Excitation and {{Measurement Patterns}} for the {{Identifiability}} of {{Directed Acyclic Graphs}}},
	year = {2022}}

@article{mapurunga-gevers-baza-DDJournal-2024,
	author = {Mapurunga, Eduardo and Gevers, Michel and Bazanella, Alexandre Sanfelice},
	journal = {IEEE Transactions on Automatic Control},
	month = {February},
	number = {2},
	pages = {1303-1308},
	title = {Identifiability of dynamical networks: the essential role of dources and dinks},
	volume = {69},
	year = {2024}}

@article{mapurunga-gevers-bazanella-necessary-2022,
	author = {Mapurunga, Eduardo and Gevers, Michel and Bazanella, Alexandre S.},
	doi = {10.1109/LCSYS.2022.3144950},
	issn = {2475-1456},
	journal = {IEEE Control Systems Letters},
	pages = {2276--2280},
	title = {Necessary and Sufficient Conditions for the Identifiability of Isolated Loops},
	year = {2022}}

@article{mapurunga-optimal-2021,
	author = {Mapurunga, Eduardo and Bazanella, Alexandre Sanfelici},
	issn = {2405-8963},
	journal = {IFAC-PapersOnLine},
	language = {en},
	month = jan,
	number = {7},
	pages = {43--48},
	series = {19th {IFAC} {Symposium} on {System} {Identification} {SYSID} 2021},
	title = {Optimal {Allocation} of {Excitation} and {Measurement} for {Identification} of {Dynamic} {Networks}},
	volume = {54},
	year = {2021}}

@article{hendrickx-gevers-bazanella-identifiability-2019,
	author = {Hendrickx, Julien M. and Gevers, Michel and Bazanella, Alexandre S.},
	doi = {10.1109/TAC.2018.2867336},
	issn = {2334-3303},
	journal = {IEEE Transactions on Automatic Control},
	month = jun,
	number = {6},
	pages = {2240--2253},
	title = {Identifiability of {{Dynamical Networks With Partial Node Measurements}}},
	volume = {64},
	year = {2019}}

@inproceedings{bazanella-gevers-hendrickx-network-2019,
	author = {Bazanella, Alexandre S. and Gevers, Michel and Hendrickx, Julien M.},
	booktitle = {2019 {{IEEE}} 58th {{Conference}} on {{Decision}} and {{Control}} ({{CDC}})},
	doi = {10.1109/CDC40024.2019.9029909},
	issn = {2576-2370},
	month = dec,
	pages = {5500--5506},
	title = {Network Identification with Partial Excitation and Measurement},
	year = {2019}}

@article{gevers-bazanella-dasilva-practical-2018,
	author = {Gevers, Michel and Bazanella, Alexandre Sanfelice and {da Silva}, Gian Vianna},
	doi = {10.1016/j.ifacol.2018.09.115},
	issn = {2405-8963},
	journal = {IFAC-PapersOnLine},
	month = jan,
	number = {15},
	pages = {862--867},
	series = {18th {{IFAC Symposium}} on {{System Identification SYSID}} 2018},
	title = {A Practical Method for the Consistent Identification of a Module in a Dynamical Network},
	volume = {51},
	year = {2018}}

@inproceedings{bazanella-gevers-hendrickx-etal-identifiability-2017,
	author = {Bazanella, Alexandre S. and Gevers, Michel and Hendrickx, Julien M. and Parraga, Adriane},
	booktitle = {2017 {{IEEE}} 56th {{Annual Conference}} on {{Decision}} and {{Control}} ({{CDC}})},
	doi = {10.1109/CDC.2017.8264547},
	month = dec,
	pages = {5870--5875},
	shorttitle = {Identifiability of Dynamical Networks},
	title = {Identifiability of Dynamical Networks: {{Which}} Nodes Need Be Measured?},
	year = {2017}}

@inproceedings{legat-hendrickx-pathbased-2021a,
	author = {Legat, Antoine and Hendrickx, Julien M.},
	booktitle = {2021 60th {{IEEE Conference}} on {{Decision}} and {{Control}} ({{CDC}})},
	doi = {10.1109/CDC45484.2021.9683062},
	issn = {2576-2370},
	month = dec,
	pages = {3024--3029},
	title = {Path-{{Based Conditions}} for {{Local Network Identifiability}}},
	year = {2021}}

@article{cheng-shi-lestas-vandenhof-DAGpartial-2025,
	author = {Cheng, Xiaodong and Shi, Shengling and Lestas, Ioannis and {Van den Hof}, Paul M. J.},
	issn = {1558-2523},
	journal = {IEEE Transactions on Automatic Control},
	month = "april",
	pages = {2305-2320},
	title = {Identifiability in Dynamic Acyclic Networks with Partial Excitation and Measurement},
	volume = {7},
	number = {4},
	year = {2025}}

@article{ramaswamy-hof-local-2021,
	author = {Ramaswamy, Karthik Raghavan and den Hof, Paul M. J. Van},
	doi = {10.1109/TAC.2020.3035634},
	issn = {1558-2523},
	journal = {IEEE Transactions on Automatic Control},
	month = nov,
	number = {11},
	pages = {5237--5252},
	title = {A {{Local Direct Method}} for {{Module Identification}} in {{Dynamic Networks With Correlated Noise}}},
	volume = {66},
	year = {2021}}

@article{cheng-shi-vandenhof-allocation-2021,
	author = {Cheng, Xiaodong and Shi, Shengling and {Van den Hof}, Paul M.J.},
	doi = {10.1109/TAC.2021.3053540},
	issn = {1558-2523},
	journal = {IEEE Transactions on Automatic Control},
	pages = {1--1},
	title = {Allocation of {{Excitation Signals}} for {{Generic Identifiability}} of {{Linear Dynamic Networks}}},
	year = {2021}}

@article{weerts-vandenhof-dankers-identifiability-2018,
	author = {Weerts, Harm H. M. and {Van den Hof}, Paul M. J. and Dankers, Arne G.},
	doi = {10.1016/j.automatica.2017.12.013},
	issn = {0005-1098},
	journal = {Automatica},
	month = mar,
	pages = {247--258},
	title = {Identifiability of Linear Dynamic Networks},
	volume = {89},
	year = {2018}}

@article{vandenhof-dankers-heuberger-etal-identification-2013,
	author = {{Van den Hof}, Paul M. J. and Dankers, Arne and Heuberger, Peter S. C. and Bombois, Xavier},
	doi = {10.1016/j.automatica.2013.07.011},
	issn = {0005-1098},
	journal = {Automatica},
	month = oct,
	number = {10},
	pages = {2994--3006},
	title = {Identification of Dynamic Models in Complex Networks with Prediction Error Methods\textemdash{{Basic}} Methods for Consistent Module Estimates},
	volume = {49},
	year = {2013}}

@inproceedings{rajagopal-ramaswamy-vandenhof-learning-2021,
	author = {Rajagopal, Venkatakrishnan C. and Ramaswamy, Karthik R. and Van Den Hof, Paul M. J.},
	booktitle = {2021 60th {{IEEE Conference}} on {{Decision}} and {{Control}} ({{CDC}})},
	doi = {10.1109/CDC45484.2021.9683377},
	issn = {2576-2370},
	month = dec,
	pages = {840--845},
	title = {Learning Local Modules in Dynamic Networks without Prior Topology Information},
	year = {2021}}

@article{van_waarde_necessary_2019,
	author = {Van Waarde, Henk J. and Tesi, Pietro and Camlibel, M. Kanat},
	journal = {IEEE Transactions on Automatic Control},
	number = 11,
	pages = {4525 -- 4537},
	title = {Necessary and {Sufficient} {Topological} {Conditions} for {Identifiability} of {Dynamical} {Networks}},
	volume = 65,
	year = {2020}}

@article{van_waarde_identifiability_2018,
	author = {van Waarde, Henk J. and Tesi, Pietro and Camlibel, M. Kanat},
	issn = {2475-1456},
	journal = {IEEE Control Systems Letters},
	language = {en},
	month = oct,
	number = {4},
	pages = {683--688},
	shorttitle = {Identifiability of {Undirected} {Dynamical} {Networks}},
	title = {Identifiability of {Undirected} {Dynamical} {Networks}: {A} {Graph}-{Theoretic} {Approach}},
	urldate = {2020-02-11},
	volume = {2},
	year = {2018}}

@book{deoGraphTheoryApplications1974,
	address = {{Englewood Cliffs, N.J}},
	author = {Deo, Narsingh},
	isbn = {978-0-13-363473-0},
	lccn = {TA338.G7 D46},
	publisher = {{Prentice-Hall}},
	series = {Prentice-{{Hall}} Series in Automatic Computation},
	title = {Graph Theory with Applications to Engineering and Computer Science},
	year = {1974}}

@string{cdc = {Proc.\ IEEE Conf.\ on Decision and Control}}

@inbook{Monika-Petreczky-Camlibel:2024,
author="J{\'o}zsa, M{\'o}nika and Petreczky, Mih{\'a}ly and Camlibel, M. Kanat",
year = {2024},
pages = {61-90},
title = {Relating the Network Graphs of State-Space Representations to Granger Causality Conditions},
booktitle = {Hybrid and Networked Dynamical Systems},
publisher={Hybrid and Networked Dynamical Systems, in Lecture Notes in Control and Information Sciences, Springer},
Editors = { R. Postoyan, P. Frasca, E. Panteley, and L. Zaccarian)},
isbn = {978-3-031-49554-0},
doi = {10.1007/978-3-031-49555-7_4}
}

@article{Granger:63,
	Author = {C.W.J. Granger},
	Journal = {Information and Control},
	Number = {1},
	Pages = {28-48},
	Title = {Economic processes involving feedback},
	Volume = {6},
	Year = {1963}}

@book{Caines:88,
	Address = {New York, USA},
	Author = {P. E. Caines},
	Publisher = {John Wiley},
	Title = {Linear stochastic systems},
	Year = {1988}}

@article{Gevers-Anderson:82,
	Author = {M. Gevers and B.D.O. Anderson},
	Journal = {IEEE Transactions on Automatic Control},
	Number = {2},
	Pages = {431-436},
	Title = {On Jointly Stationary Feedback Free Stochastic Processes},
	Volume = {27},
	Year = {1982}}

\end{document}